\documentstyle[12pt,epsfig]{article}
\input{epsf}
\newcommand{\be}{\begin{equation}}
\newcommand{\ee}{\end{equation}}
\newcommand{\bea}{\begin{eqnarray}}
\newcommand{\eea}{\end{eqnarray}}

\newcommand{\lam}{\lambda}
\newcommand{\bQ}{\bf Q}
\newcommand{\bk}{\bf k}

\def\la{\mathrel{\mathpalette\fun <}}
\def\ga{\mathrel{\mathpalette\fun >}}
\def\fun#1#2{\lower3.6pt\vbox{\baselineskip0pt\lineskip.9pt
\ialign{$\mathsurround=0pt#1\hfil##\hfil$\crcr#2\crcr\sim\crcr}}}

\begin{document}

\title{Two-photon partial widths of
 tensor mesons }
\author{A.V. Anisovich, V.V. Anisovich,
M.A. Matveev, and V.A. Nikonov}
\date{}
\maketitle

\begin{abstract}

We calculate partial widths of the $\gamma\gamma$ decay of
the tensor $q\bar q$ states
$a_2(1320)$, $f_2(1270)$, $f_2(1525)$,
their radial excitations
$a_2(1660)$, $f_2(1640)$, $f_2(1800)$
as well as $^3F_2 q\bar q$ states.
Calculations are performed in the framework of the same approach which
was used before for the study of  radiative decays $f_0(980)\to
\gamma\gamma$, $a_0(980)\to \gamma\gamma$ and $\phi(1020)\to \gamma
f_0(980)$: the assumption made is that of $q\bar q$ structure of
$f_0(980)$ and $a_0(980)$ [A.V. Anisovich et al., Phys. Lett. B {\bf
456}, 80 (1999); Eur. Phys. J. A {\bf 12}, 103 (2001)]. The
description of the decay partial widths for $a_2(1320)$, $f_2(1270)$,
$f_2(1525)$ and $f_0(980)$, $a_0(980)$ is reached with the
approximately equal radial wave functions, thus giving a strong
argument in favour of the
 fact that these scalar  and tensor mesons are to be classified as
 members of the same $P$-wave $q\bar q$ multiplet.

\end{abstract}

\section{Introduction}

For the time being the main problem of meson spectroscopy is
the reliable determination of states belonging to the $P$-wave
$q\bar q$ multiplet $1^3P_Jq\bar q$. The solution of this problem
is of principal importance for the quark systematics as well as for the
 search of exotic mesons such as glueballs and hybrids (in connection
with this problem see [1-4]).
%\cite{klempt}--\cite{ufn}).
The classification of mesons $f_0(980)$ and $a_0(980)$, that is of crucial
meaning
for the nonet of scalar mesons $1^3P_0 q\bar q$, gives rise to
certain questions. In a set of papers [5-10],
%\cite{km1900}--%,kaon,radius,f(980),f0gg,\cite{phi},
on the basis of the
analysis of experimental data, it was argued that for the states
$f_0(980)$ and $a_0(980)$ the $1^3P_0 q\bar q$ component is dominant.
However, another point of view on the structure of these mesons  exists
as well, see mini-review \cite{ST} and references therein.

The investigation of radiative decays is a powerful tool for
establishing the quark structure of hadrons.  At early stage of the
quark model,  radiative decays of vector mesons provided strong
evidence for  constituent quark being a universal constructive element
of mesons and baryons [12-15].
%\cite{V-gP1}--,V-gP2,V-gP3,\cite{V-gP4}.
In our opinion,
the radiative decays of the $1^3P_J q\bar q$ mesons are equally
important for the determination of the $P$-wave multiplet.

Partial widths of the decays $f_0(980)\to \gamma\gamma$
and $a_0(980)\to \gamma\gamma$ have been calculated in \cite{f0gg}
assuming the mesons
$f_0(980)$ and $a_0(980)$ to be dominantly  $q\bar q$ states, that
is, $1^3P_0q\bar q$ mesons. The results of our calculation agree
well with experimental data. In paper \cite{phi}, on the basis of
data \cite{novosib} for the decay
$\phi(1020)\to \gamma f_0(980)$ together with the
value of partial width $f_0(980)\to \gamma\gamma$
obtained in the re-analysis \cite{pennington}, the
flavour content of $f_0(980)$ has been studied. Assuming the flavour
wave function in the form $n\bar n \cos \varphi+s\bar s \sin \varphi$,
we have described the experimental data with two allowed values of
mixing angle: either $\varphi=-48^\circ\pm 6^\circ$ or
$\varphi=85^\circ\pm 4^\circ$ (the negative angle is more preferable).
Both values of mixing angle are in qualitative agreement with data on
hadronic decays of $f_0(980)$ into $\pi\pi$ and $K\bar K$
\cite{f(980),ANS}.

Although direct calculations  of widths of radiative
decays agree well with the hypothesis that the $q\bar q$ component
dominates both $f_0(980)$ and $a_0(980)$, to be confident that these
mesons are members of $1^3P_0q\bar q$ multiplet one more step is
necessary:  it is necessary to check whether radiative decays of tensor
mesons $a_2(1320)$, $f_2(1270)$,  $f_2(1525)$ can be calculated under
the same assumption and within the same technique as it was done for
the reactions involving $f_0(980)$ and $a_0(980)$. The tensor mesons
$a_2(1320)$, $f_2(1270)$,  $f_2(1525)$ are basic members of the
$P$-wave $q\bar q$ multiplet, and just the existence of  tensor
mesons lays in the ground of the nonet classification of  mesons as
$q\bar q$-states, with four $P$-wave nonets \cite{Pqq1,Pqq2}.

In the framework of spectral integration technique, we calculate
the transition form factors of tensor mesons $a_2(1320)
\to \gamma^*(Q^2)\gamma$, $f_2(1270) \to \gamma^*(Q^2)\gamma$ and
$f_2(1525) \to \gamma^*(Q^2)\gamma$
in the region of small $Q^2$:
these form
factors, in the limit $Q^2 \to 0$, determine partial widths
$a_2(1320)\to \gamma\gamma $,
$f_2(1270) \to \gamma\gamma$ and $f_2(1525) \to \gamma\gamma$.
The spectral representation technique
has been developed in \cite{eta}
for the investigation of the
transitions of pseudoscalar mesons such as
$\pi^0 \to \gamma^*(Q^2)\gamma$, $\eta \to
\gamma^*(Q^2)\gamma$ and $\eta' \to \gamma^*(Q^2)\gamma$.
As is said above, by using this technique the calculation of the decay
coupling constants  $f_0(980)\to \gamma\gamma$,
$a_0(980)\to \gamma\gamma$ and $\phi(1020)\to \gamma f_0(980)$ has been
performed in \cite{f0gg,phi}.

In the region of moderately small $Q^2$, where Strong-QCD works, the
transition form factor $q\bar q$-meson $\to \gamma^*(Q^2)\gamma$ is
determined by the quark loop diagram of Fig. 1$a$ which is a convolution
of the $q\bar q$-meson and photon wave functions, $\Psi_{q\bar q}
\otimes \Psi_{\gamma}$.  The calculation of the process of Fig. 1$a$ is
performed in terms of the double spectral representation over $q\bar
q$ invariant masses squared, $s=(m^2+k^2_{\perp})/\left ( x(1-x)\right
)$  and $s'=(m^2+k'^2_{\perp})/\left ( x(1-x)\right )$ where
$k^2_{\perp}$, $k'^2_{\perp}$ and $x$ are the light-cone variables and
$m$ is the constituent quark mass. Following \cite{eta}, we represent
photon wave function as a sum of the two components which describe the
prompt production of the $q\bar q$ pair at large $s'$ (with a
point-like vertex for the transition $\gamma \to q\bar q $,
correspondingly) and the production in the low-$s'$ region where the
vertex $\gamma \to q\bar q $ has a nontrivial structure due to soft
$q\bar q$ interactions.  The necessity to include such a component can
be argued, for example, by the vector-dominance model
$\gamma\to\rho^0,\omega,\phi\to q\bar q$.

The process of Fig. 1$a$ at moderately small
$Q^2$ is mainly determined by the low-$s'$ region, in other words by
the soft component of photon wave function.

The soft component of the photon wave function was restored in
\cite{eta}, on the basis of the experimental data for the transition
$\pi^0 \to \gamma^*(Q^2)\gamma$ at $Q^2 \leq 1$ GeV$^2$.  Once the
photon wave function is found, the form factors $a_2 \to
\gamma^*(Q^2)\gamma$ and $f_2 \to \gamma^*(Q^2)\gamma$ at $Q^2 \leq
1$ GeV$^2$ provide us the opportunity to investigate in detail the
tensor-meson wave functions.  However, when investigating a
small-$Q^2$ region, we
may restrict ourselves with a simplified,
one-parameter wave function of the basic tensor
mesons $1^3P_2q\bar q$, this parameter being  the mean radius squared
$R^2_T$.

Assuming the $q\bar q$ structure of tensor
mesons, the flavour content of $a_2(1320)$ is fixed, thus allowing
unambiguous calculation of the transition form factor $a_2(1320) \to
\gamma\gamma$. Reasonable agreement with data has been obtained at
$7\,{\rm GeV}^{-2}\le R^2_{a_2(1320)} \le 12\,{\rm GeV}^{-2}$
(remind that for pion
$R^2_{\pi} \simeq 10$ GeV$^{-2} \simeq 0.4$ fm$^2$). To describe the
decay $a_0(980)\to \gamma\gamma$ the quark wave function of $a_0(980)$
should have nearly the same mean radius square \cite{f0gg}
$7\,{\rm GeV}^{-2}\le R^2_{a_0(980)} \le 12\,{\rm GeV}^{-2}$. Still, we
do not exclude the possibility that the $P$-wave states may be rather
compact. Hadronic reactions agree with this possibility:
the estimation of radius of $f_0(980)$  carried out
by using GAMS data for $\pi^- p\to \pi^0\pi^0n$
\cite{GAMS} proves
that $q\bar q$ component in $f_0(980)$ gives
 \cite{radius}:
$ R^2_{f_0(980)}=(6\pm 6)$ GeV$^{-2}$.

Partial widths $\Gamma (f_2(1270) \to  \gamma\gamma)$ and
$\Gamma (f_2(1525) \to  \gamma\gamma)$ depend on the
relative weights of strange and non-strange components in a
tensor-isoscalar meson, $s\bar s$ and $n\bar n$.
The study of hadronic decays tells us that $f_2(1270)$
is dominantly $n\bar n$ state, while $f_2(1525)$ is, correspondingly,
$s\bar s$ one. It is in accordance with the calculated values of
partial widths $\Gamma (f_2(1270) \to  \gamma\gamma)$ and
$\Gamma (f_2(1525) \to  \gamma\gamma)$: at
$R^2_{f_2(1270)}\simeq R^2_{f_2(1525)} \sim R^2_{f_0(980)}$,
the agreement with  data is reached with
$n\bar n$- and $s\bar s$-dominated components in $f_2(1270)$ and
$f_2(1525)$, respectively.

The two-photon decays of radial-excited states, $2^3P_2q\bar q\to
\gamma\gamma$, are suppressed as compared to decays of basic states.
The reason is that radial wave functions of the states $2^3P_2q\bar q$
change sign, so the convolution of wave function
$\psi_{2^3P_2q\bar q}\otimes \psi_{\gamma}$ is comparatively small.
This fact is also the reason of a qualitative character of predictions
for the decays $2^3P_2q\bar q \to \gamma\gamma$.

The paper is organized as follows.
In Section 2 we present basic formulae for the calculation of
 amplitudes for
$\gamma\gamma$ decaying into tensor mesons,
members of the $1^3P_2 q\bar q$ and $2^3P_2 q\bar q$ nonets.
The results of the calculation are given in Section 3. In Conclusion
we discuss the $q\bar q$ multiplet
classification of tensor and scalar mesons resulting
from radiative meson decays.

\section{Tensor-meson decay amplitudes for \\ the process
$q\bar q\; (2^{++})\to \gamma\gamma$ }

Below the formulae are presented for the amplitudes
of radiative decay of the  $q\bar q$ tensor mesons
belonging to the multiplets located at $\la 2000$ MeV:
$1^3P_2 q\bar q$, $2^3P_2 q\bar q$ and $1^3F_2 q\bar q$.

\subsection{Spin-momentum structure of the decay amplitude}

The decay amplitude for the process
$q\bar q\; (2^{++})\to \gamma\gamma$ has
the following structure:
\be
A^{(T)}_{\mu\nu,\alpha\beta}= e^2 \left [
S^{(0)}_{\mu\nu,\alpha\beta} (p,q)
 \; F^{(0)}_{T\to \gamma\gamma}(0,0) +
S^{(2)}_{\mu\nu,\alpha\beta} (p,q)
 \; F^{(2)}_{T\to \gamma\gamma}(0,0)
\right ]\, ,
\label{2.1}
\ee
where $e$ is the electron charge ($e^2 /4\pi =\alpha = 1/137$).
Here $S^{(0)}_{\mu\nu,\alpha\beta}$ and
$S^{(2)}_{\mu\nu,\alpha\beta}$ are the moment-operators, indices
$\alpha,\beta$ refer to photons and $\mu,\nu$ to  tensor meson.
The transition form factors for  photons
with  transverse polarization
$T \to \gamma_\perp (q_1^2)\gamma_\perp (q_2^2)$,
namely, $ F^{(0)}_{T\to \gamma\gamma}(q_1^2 ,q_2^2)$ and
$ F^{(2)}_{T\to \gamma\gamma}(q_1^2 ,q_2^2)$,
depend on the photon momenta squared $q_1^2 $ and $q_2^2 $;
the limit values $q_1^2 =0$ and $q_2^2 =0$  correspond to the
two-photon decay. We also use the notations: $p=q_1+q_2$ and $q
=(q_1-q_2)/2$.

The moment-operators read:
\be
S^{(0)}_{\mu\nu,\alpha\beta}(p,q)=
g^{\perp\perp}_{\alpha\beta}
\left ( \frac{q_\mu q_\nu}{q^2}-\frac13
g^\perp_{\mu\nu} \right )
\label{2.2}
\ee
and
\be
 S^{(2)}_{\alpha \beta \; ,\; \mu\nu}(p,q)
=
g^{\perp\perp}_{\mu\alpha} g^{\perp\perp}_{\nu\beta} +
g^{\perp\perp}_{\mu\beta} g^{\perp\perp}_{\nu\alpha}
-g^{\perp\perp}_{\mu\nu}
g^{\perp\perp}_{\alpha\beta} \ ,
\label{2.3}
\ee
where
metric tensors $ g_{\mu\nu}^{\perp}$ and
$g_{\alpha\beta}^{\perp\perp}$ are determined as follows:
\be
g^{\perp}_{\mu\nu}=g_{\mu\nu}-
\frac{p_\mu p_\nu}{p^2} ,\qquad
g^{\perp\perp}_{\alpha\beta}=g_{\alpha\beta}-
\frac{q_\alpha q_\beta}{q^2} -
\frac{p_\alpha p_\beta}{p^2}.
\label{2.4}
\ee
The moment-operators are orthogonal in the space of photon
polarizations:\\
$
S^{(0)}_{\mu\nu,\alpha\beta}
S^{(2)}_{\mu'\nu',\alpha\beta}=0$.

The spin structure of the amplitude
$A^{(T)}_{\mu\nu,\alpha\beta}$ is discussed in more detail in
Appendix A, also there is presented a connection between amplitudes
written in terms of spin operators and standard helicity amplitudes.

\subsection{Form factor $F^{(H)}_{T\to \gamma\gamma}(q_1^2,q_2^2)$}

Following the prescription of \cite{eta},
we write down the amplitude of the process of Fig. 1$a$ in terms of the
spectral representation in the channels related to tensor meson and
photon $\gamma (q_2)$.
The double spectral representation for the form factor
$F^{(H)}_{T\to \gamma\gamma}(q_1^2,q_2^2)$ with $H=0,2$ reads:
\be
F^{(H)}_{T\to \gamma\gamma}(q^2_1,q^2_2) =
Z_T \sqrt {N_c} \int \limits_{4m^2}^{\infty}
\frac {ds\;ds'}{\pi^2}  \frac {G_{T\to q\bar q}(s)}{s-m_{T}^2}\times
\label{2.5}
\ee
$$
\times  d\Phi(P,P';k_1,k'_1,k_2) S^{(H)}
(P\; ^2,P\; '^2,\tilde q^2)
\;     \frac {G_{\gamma \to q\bar q}(s')}{s'-q_2^2}\,.
$$
In the spectral integral (\ref{2.5}),
the momenta of the intermediate states differ from those of
the initial/final states. The corresponding momenta for
intermediate states are
re-denoted as is shown in Fig. 1$b$:
\be
q_1 \to P\;-P', \qquad q_2 \to P', \qquad p\to P \; ,
\label{2.6}
\ee
$$
P^2=s, \qquad P\;'^2=s', \qquad  ( P\;'-P)^2=\tilde q^2=q^2_1.
$$
It should be stressed that we fix $\tilde q^2=q^2_1$, although
$ P\;'-P=\tilde q \ne q_1$.
The triangle-diagram phase space
$ d\Phi(P,P';k_1,k'_1,k_2)$ is equal to:
\be
d\Phi(P,P';k_1,k'_1,k_2)=
\frac 14 \; \frac {d^3k_1}{(2\pi)^3 2k_{10}}
\frac {d^3k'_1}{(2\pi)^3 2k'_{10}}
\frac {d^3k_2}{(2\pi)^3 2k_{20}}\times
\label{2.7}
\ee
$$
\times (2\pi)^4\delta^{(4)} \left (P -k_1-k_2 \right )
(2\pi)^4\delta^{(4)} \left (P\; ' -k'_1-k_2 \right ).
$$
The factor $Z_{T}$ is determined by the quark content of tensor
meson. For the $a_2$-meson and $n\bar n$ or $s\bar s$ components in
the $f_2$-meson, the charge factors are equal to
\be
Z_{a_2}=2\frac{e_u^2-e_d^2}{\sqrt 2 }\; , \qquad
Z_{n\bar n}=2\frac{e_u^2+e_d^2}{\sqrt 2 } \; ,\qquad
Z_{s\bar s}=2e_s^2  \, .
\label{2.8}
\ee
The factor $\sqrt {N_c}$, where
$N_c=3$ is the number of colours, is related to the normalization of
the photon vertex made in \cite{eta}.
We have two sorts of diagrams: with quark lines drawn clockwise and
anticlockwise; the factor $2$ in
(\ref{2.8}) stands for this doubling.
The vertices
$G_{\gamma \to n\bar n}(s')$ and $G_{\gamma \to s\bar s}(s')$ were
found in \cite{eta}; the photon wave function
$G_{\gamma \to n\bar n}(s)/s$ is shown in Fig. 2.

\subsubsection{Wave functions for the $1^3P_2 q\bar q$ and
$2^3P_2 q\bar q$ states}

We parametrize the wave functions of mesons of the
basic multiplet, $1^3P_2 q\bar q$, in the exponential
form:
\be
\Psi_{T}(s)=\frac {G_{T}(s)}{s-m_{T}^2} = Ce^{-bs},
\label{2.9}
\ee
where $C$ is the normalization constant,
$\Psi_{T}\otimes \Psi_{T}=1$, and the parameter $b$ can
be related to the tensor-meson radius squared.

For mesons
of the first radial excitation, $2^3P_2 q\bar q$,
the wave functions can be written using exponential approximation
as
\be
\Psi_{T1}(s)= C_1e^{-b_1s} (D_1 s-1).
\label{2.9a}
\ee
The parameter $b_1$ can be related to the radius of the radial
excitation state, then the values $C_1$ and $D_1$ are fixed by
the normalization and orthogonality requirements,
$\Psi_{T1}\otimes \Psi_{T1}=1$ and $\Psi_{T}\otimes \Psi_{T1}=0$.

\subsection{Spin structure factors $S^{(0)}
(P\; ^2,P\; '^2,\tilde q^2)$
and \\ $S^{(2)}(P\; ^2,P\; '^2,\tilde q^2)$}

For the amplitude of Fig. 1$b$ with transverse polarized photons,
the spin structure factors
are fixed by the vertex for transition $T\to q\bar q$, and
we denote this vertex as $T_{\mu\nu} $. One has:
\be
\mathrm{Tr} \left [\gamma^{\perp\perp}_{\beta} (\hat k'_1+m)
\gamma^{\perp\perp}_{\alpha} (\hat k_1+m)T_{\mu\nu} (\hat k_2-m)
\right ] =
\label{2.10}
\ee
$$
=S^{(0)}_{\mu\nu,\alpha\beta} (P,P'_\perp)
S^{(0)}(P\; ^2,P\; '^2,\tilde q^2)+
S^{(2)}_{\mu\nu,\alpha\beta} (P,P'_\perp)
S^{(2)}(P\; ^2,P\; '^2,\tilde q^2)
$$
Here $\gamma^{\perp\perp}_{\alpha}$
and $\gamma^{\perp\perp}_{\beta}$ stand for photon vertices,
$ \gamma^{\perp\perp}_{\alpha} =
g_{\alpha\alpha '}^{\perp\perp}\gamma_{\alpha '}$,
while $g_{\alpha\alpha '}^{\perp\perp}$ is determined by (\ref{2.4})
with the substitution
$q_1 \to P\;-P'$ and $ q_2 \to P'$. The moment-operators
$S^{(0)}_{\mu\nu,\alpha\beta} (P,P'_\perp)$ and
$S^{(2)}_{\mu\nu,\alpha\beta} (P,P'_\perp)$ work also in the
intermediate-state momentum space.
Recall that the momenta $k'_1$, $k_1$
and $k_2$ in (\ref{2.10}) are mass-on-shell.

The vertex $T_{\mu\nu}$ taken in a minimal form reads:
\be
T_{\mu\nu}^{(k)} = k_\mu \gamma_\nu + k_\nu \gamma_\mu -\frac 23
g_{\mu\nu}^\perp \hat k \ ,
\label{2.11}
\ee
where $k=k_1-k_2$ and $g_{\mu\nu}^\perp P_\nu =0$.
With $T_{\mu\nu}$ determined by (\ref{2.11}), we present
the spin structure factors
$S^{(H)}(P\; ^2,P\; '^2,\tilde q^2)$
at $q_2^2 =0$ and small $q_1^2 = \tilde q^2 \equiv -Q^2$.
Below we denote $\Sigma =(s+s')/2$ and $\Delta =s-s' $ and
take into account that $\Delta \sim Q$. For
the non-vanishing terms
in the limit $Q^2\to 0$, we have:
\be
S^{(0)}(P\; ^2,P\; '^2,-Q^2)=
\frac{64m^2\Sigma^2Q^4}{(\Delta^2+4\Sigma
Q^2)^2}\left(4m^2-\Sigma\right)+
\label{2.12}
\ee
$$ +\frac{4\Sigma
Q^2\Delta^2}{(\Delta^2+4\Sigma Q^2)^2}
\left(32m^4+8m^2\Sigma-3\Sigma^2\right)+
\frac{4m^2\Delta^4}{(\Delta^2+4\Sigma Q^2)^2}\left(4m^2+3\Sigma\right)
$$
and
\be
S^{(2)}(P\; ^2,P\; '^2,-Q^2)=
\frac{8\Sigma^2Q^4}{(\Delta^2+4\Sigma
Q^2)^2}\left(-16m^4+\Sigma^2\right)+
\label{2.13}
\ee
$$
+\frac{4\Sigma Q^2\Delta^2}{(\Delta^2+4\Sigma
Q^2)^2}\left(-16m^4-4m^2\Sigma+\Sigma^2\right)+
\frac{4m^2\Delta^4}{(\Delta^2+4\Sigma Q^2)^2}\left(-2m^2-\Sigma\right).
$$

\subsubsection{Spin structure factors $S^{(H)}
(P\; ^2,P\; '^2,\tilde q^2)$ for
pure $q\bar q\, (L=1)$ and $q\bar q\; (L=3)$ states}

The $q\bar q\; (2^{++})$  state can be constructed in two ways, namely,
with the $q\bar q$ orbital momenta $L=1$ and $L=3$
(the $^3P_2q\bar q$ and $^3F_2q\bar q$ states). The vertex
$T_{\mu\nu}$ of Eq. (\ref{2.11}), corresponding to the dominant
$P$-wave $q\bar q$ state, includes also certain admixture of the
$F$-wave $q\bar q$ state.

The vertex
for the production of pure $q\bar q\; (L=1)$ state reads:
\be
T^{(L=1),(k)}_{\mu\nu} = k_\mu \Gamma_\nu + k_\nu \Gamma_\mu -\frac 23
g_{\mu\nu}^\perp (\Gamma k), \qquad
\Gamma_\mu =\gamma ^\perp _\mu -\frac{k_\mu}{2m+\sqrt s}\; ,
\label{2.14}
\ee
where the operator $\Gamma_\mu $ selects the spin-1 state for
$q\bar q$ (see \cite{spin-1,spin-2} for details). We present
corresponding spin factors,
$S_{L=1}^{(0)}(P\; ^2,P\; '^2,-Q^2)$ and
$S_{L=1}^{(2)}(P\; ^2,P\; '^2,-Q^2)$, in Appendix B.

The $(L=3)$-operator for the $^3F_2q\bar q$ state is equal to:
\be
T^{(L=3),(k)}_{\mu\nu} = k_\mu k_\nu (\Gamma k) -\frac {k^2}{5}
\left (
g_{\mu\nu}^\perp (\Gamma k) +
\Gamma_\mu k_\nu + \Gamma_\nu k_\mu \right ).
\label{2.15}
\ee
Corresponding spin factors,
$S_{L=3}^{(0)}(P\; ^2,P\; '^2,-Q^2)$ and
$S_{L=3}^{(2)}(P\; ^2,P\; '^2,-Q^2)$, are also given in Appendix B.

\subsection{ Spectral integral representation}

In formula (\ref{2.5}) one
can integrate over the phase space by  using $\delta$-functions,
with fixed the energies squared, $s$ and $s'$. So we have:
\be
F^{(H)}_{T\to \gamma\gamma}(-Q^2,0) =
\label{2.17}
\ee
$$
=Z_T \sqrt {N_c}
\int \limits_{4m^2}^\infty
\frac{dsds'}{\pi^2} \psi_T(s)\psi_{\gamma}(s')
\frac {\theta\left
(ss'Q^2-m^2\lambda(s,s',-Q^2)\right )}{16\sqrt{\lambda(s,s',-Q^2)}}
S^{(H)}_{T \to\gamma \gamma}(s,s',-Q^2)\ ,
$$
with
\be
\lambda(s,s',-Q^2)= (s'-s)^2+2Q^2(s'+s)+Q^4\ .
\label{2.18}
\ee
The $\theta$-function restricts the integration region
for different $Q^2$: $\theta(X)=1$ at $X\ge 0$ and $\theta(X)=0$
at $X < 0 $.

In the limit $Q^2 \to 0$, one has
$$
F^{(H)}_{T\to \gamma \gamma}(-Q^2\to 0,0)=
Z_T \sqrt {N_c}
\int \limits_{4m^2}^\infty
\frac{d\Sigma }{\pi} \psi_T(\Sigma)\psi_{\gamma}(\Sigma)
\int
\limits_{-b}^
{+b}
\frac{ d\Delta}{\pi}\;  \frac {
S^{(H)}_{T \to\gamma \gamma}(s,s',-Q^2)}
{16\sqrt{\Lambda(\Sigma ,\Delta ,Q^2)}}
$$
\be
b=Q\sqrt{\Sigma (\Sigma /m^2 -4)}, \qquad
\Lambda(\Sigma ,\Delta, Q^2)=\Delta^2+4\Sigma Q^2 \ .
\label{2.19}
\ee
Where the spin factors
$ S^{(H)}_{T \to\gamma \gamma}(s,s',-Q^2) $ are given in
(\ref{2.12}) and (\ref{2.13}).

The integration over $\Delta$ performed, the spectral
representation
for the form factor
$F^{(H)}_{T\to \gamma \gamma}(0,0)$ reads:
\be
F^{(H)}_{T\to \gamma \gamma}(0,0)=
\frac{Z^{(q\bar q)}_T\sqrt{N_c}}{16\pi}
\int \limits_{4m^2}^\infty
\frac{ds}{\pi} \psi_T(s)\psi_{\gamma}(s)I^{(H)}(s)
\label{2.20}
\ee
where
\be
I^{(0)}(s)=-2\sqrt{s\left(s-4m^2\right)}
\left(12m^2+s\right)
+4m^2\left(4m^2+3s\right)
\ln\frac{s+\sqrt{s\left(s-4m^2\right)}}
        {s-\sqrt{s\left(s-4m^2\right)}}
\label{2.21}
\ee
and
\be
I^{(2)}(s)=\frac{4\sqrt{s\left(s-4m^2\right)}}{3}
\left(5m^2+s\right)
-4m^2\left(2m^2+s\right)
\ln\frac{s+\sqrt{s\left(s-4m^2\right)}}
        {s-\sqrt{s\left(s-4m^2\right)}}
\label{2.22}
\ee
The tensor-meson decay
form factors $T\to \gamma\gamma$ with vertices $T\to q\bar q$ for
pure $q\bar q\, (L=1)$ and $q\bar q\; (L=3)$ states
(see (\ref{2.14}) and (\ref{2.15})) are given in Appendix C.

\subsection{Light-cone variables}

The  formula (\ref{2.5}) allows one to make easily
the  transformation to
the light-cone variables using the boost along the $z$-axis.
Let us use the frame
in which  initial  tensor meson is moving along the
z-axis with the momentum
$p\to \infty$:
\be
P=(p+\frac {s}{2p}, 0, p), \qquad
P\; '=(p+\frac {s'+Q^2}{2p},\mathbf{Q}, p).
\label{2.23}
\ee
Then the transition form factor $T \to \gamma^*(Q^2) \gamma  $
reads:
\be
F^{(H)}_{T\to \gamma\gamma}(-Q^2,0)=
\frac {Z_T \sqrt {N_c} }{16\pi^3}
\int \limits_{0}^{1}
\frac {dx}{x(1-x)^2}  \int d^2k_{\perp} \Psi_T (s) \Psi_{\gamma}(s')
S^{(H)}(s,s',-Q^2)\ ,
\label{2.24}
\ee
where $ x=k_{2z}/p$, $ \mathbf{k}_{\perp}= \mathbf{k}_{2\perp}$, and
the $q\bar q$ invariant mass squares are
\be
s=\frac{m^2+k^2_{\perp} }{x(1-x)}, \qquad
s'=\frac{m^2+(-  x \bQ +\bk_\perp )^2  }{x(1-x)} .
\label{2.25}
\ee

\subsection{Tensor-meson charge form factor}

In order to relate the   wave function parameters $C$ and $b$ entering
(\ref{2.9})
to the tensor-meson radius squared, we calculate the
meson charge form factor
averaged over polarizations; corresponding process is shown
diagrammatically in Fig. 1$c$.
Thus determined  form factor amplitude has the structure as follows:
\be
A_{\mu}=(p_\mu+ p'_\mu)F_T(-Q^2) \ ,
\label{2.26}
\ee
where the meson charge form factor $F_T(Q^2)$ is the convolution of
the tensor-meson wave functions $\Psi_T\otimes\Psi_T$.

\subsubsection{Charge form factor in the light-cone variables}

Using light-cone variables, one
can express the $q\bar q $-meson charge form factor as follows
(for example, see \cite{phi,eta}):
\be
F_T(-Q^2 ) = \frac {1}{16\pi^3}
\int \limits_{0}^{1}
\frac {dx}{x(1-x)^2}  \int d^2k_{\perp} \Psi_T (s) \Psi_T (s')
S_T(s,s',-Q^2)\ ,
\label{2.27}
\ee
$S_T(s,s',-Q^2)$ being determined by the quark
loop trace in the intermediate state:
$$
\frac 15 \mathrm{Tr} [T_{\mu\nu}^{(k)} (\hat k_1+m)
\gamma_{\alpha} (\hat k'_1+m)T_{\mu\nu}^{(k')}  (\hat k_2-m)] =
$$
\be
=[ P'_{\alpha} +P_{\alpha} -
\frac{ s'-s   }{Q^2 }  (  P'_{\alpha} -P_{\alpha} ) ]
\;S_T(s,s',-Q^2),
\label{2.28}
\ee
where $(  P' -P)^2 =-Q^2$.

To relate the wave function parameters to the tensor-meson radius
squared we may restrict ourselves by the consideration of the low-$Q^2$
region.  The low-$Q^2$  charge form factor can be expanded in a series
over $Q^2$:
\be
F_T(-Q^2 ) \simeq 1-\frac{1}{6}R_T^2 Q^2\ .
\label{2.29}
\ee
At small $Q^2$ one has for $S_T(s,s',-Q^2)$:
\be
\label{2.30}
S_T(s,s',-Q^2)=
\frac{Q^2}{45(\Delta^2+4\Sigma Q^2)\Sigma^2}\left [
-48\Sigma^3\left (8m^2+3\Sigma\right )\left( 4m^2-\Sigma\right )-
\right  .
\ee
$$
\left .
-16Q^2\Sigma^2\left(64m^4-22m^2\Sigma+9\Sigma^2\right)
+8\Sigma\Delta^2\left(40m^4+29m^2\Sigma-9\Sigma^2\right)\right]-
$$
$$
-\frac{Q^2\Sigma^3\left(8m^2+3\Sigma\right)\left(4m^2-\Sigma\right)}
{45(\Delta^2+4\Sigma Q^2)^2}
\left[\frac{\Delta^4}{4\Sigma^4}+\frac{Q^2\Delta^2}{\Sigma^3}
-\frac{Q^4}{\Sigma^2}\right]\ .
$$
Recall, $\Sigma =(s+s')/2$ and $\Delta =s-s'$.

\subsubsection{Spectral integral representation
for charge form factor}

Expanded in a series over $Q^2$, the spectral integral for
charge form factor, $F_T(-Q^2 )$,
reads:
\be
F_{T}(-Q^2)\simeq \frac 1{16\;\pi}
\int \limits_{4m^2}^\infty
\frac{ds}{\pi} \psi^2_T(s)
\left ( I(s) - \frac{Q^2}{6} I_R(s) \right )\ ,
\label{2.31}
\ee
where
\be
I(s)=\frac{8\sqrt{s\left (s-4m^2\right)}}{15s}
\left(8m^2+3s\right)\left(s-4m^2\right )\ ,
\label{2.32}
\ee
$$
I_R(s)=\frac{4\sqrt{s\left(s-4m^2\right)}}{5s^3}
\left(-64m^6-40m^4s+26m^2s^2+9s^3\right)+
$$
$$
+\frac{8}{15s}\left(16m^4-46m^2s+9s^2\right)
\ln\frac{s+\sqrt{s\left(s-4m^2\right)}}
        {s-\sqrt{s\left(s-4m^2\right)}} .
$$
Comparison of this expression with formula (\ref{2.29})
gives us the parameters of the tensor-meson wave function.

Formulae (\ref{2.31}) and (\ref{2.32}) can be used for the
determination of the wave function parameters for any $n^3P_2q\bar q$
meson.

\section{Results}

Here are presented the results of the
calculation of the $\gamma\gamma$ partial widths for
mesons of the basic multiplet $1^3P_2q\bar q$, that is, $a_2(1320)$,
$f_2(1270)$ and $f_2(1525)$. According to \cite{syst}, mesons
of the first radial excitation, $2^3P_2q\bar q$, are
 $a_2(1660)$, $f_2(1640)$ and $f_2(1800)$), and we calculate their
$\gamma\gamma$ partial widths  as well.
We estimate also the $\gamma\gamma$ widths of the $F$-wave mesons,
namely, the members of $1^3F_2q\bar q$ nonet, these mesons are located
in the vicinity of $2000$ MeV \cite{syst}.

\subsection{Tensor-meson $\gamma\gamma$ decay partial width}

Partial width, $\Gamma_{T \to \gamma \gamma}$, is determined as
follows:
\bea
m_{T}\Gamma_{T \to \gamma \gamma} &=&
\frac 12  \int d\Phi_2(p;q_1,q_2)
\frac 15
\sum_{\mu\nu,\alpha\beta} |A_{\mu\nu,\alpha\beta}|^2=
\\
&=& \frac 45
\pi \alpha^2\left [ \frac 13
\left ( F^{(0)}_{T\to \gamma\gamma}(0,0)\right )^2 +
  \left (F^{(2)}_{T\to \gamma\gamma}(0,0)\right )^2 \right ]\ .
\nonumber \label{2.33}
\eea
Here $m_T$ is the  mass of the tensor meson;
the summation is carried out over outgoing-photon
polarizations; the photon identity factor $ 1/2$  is written
explicitly; averaging over the tensor-meson polarizations results in
the factor $1/5$. The two-particle invariant phase space is equal to
\be
d\Phi_2(p;q_1,q_2) = \frac 12 \; \frac {d^3q_1}{(2\pi)^3 2q_{10}}
\; \frac {d^3q_2}{(2\pi)^3 2q_{20}}\; (2\pi)^4\delta^{(4)} \left (
p -q_1-q_2 \right ),
\label{2.34}
\ee
and for photons $\int d\Phi_2(p;q_1,q_2) = 1/16\pi$.

\subsection{Transition form factors $F^{(H)}_{n\bar n\to\gamma\gamma}$
and $F^{(H)}_{s\bar s\to\gamma\gamma}$ for mesons
of the $1^3P_2 q\bar q$ nonet}

The transition form factors $F^{(H)}_{n\bar n\to\gamma\gamma}$ and
$F^{(H)}_{s\bar s\to\gamma\gamma}$ are determined by Eq.
(\ref{2.20}). They depend on the quark mass and type of the vertex
entering the spin factor, as well as the
tensor mesom wave function. For
non-strange quarks, the ratios
$F^{(H)}_{a_2\to\gamma\gamma}(0,0)/Z_{a_2}$ and $F^{(H)}_{n\bar
n\to\gamma\gamma}(0,0)/Z_{n\bar n}$ are equal to one another, provided
$a_2$ and $(n\bar n)_2$-state belong to the same $q\bar q$ multiplet.
The magnitudes $F^{(H)}_{n\bar n\to\gamma\gamma}(0,0)$ and
$F^{(H)}_{s\bar s\to\gamma\gamma}(0,0)$ are shown in Fig. 3 for
different tensor mesons.

The transition form factors for mesons of the basic multiplet
$1^3P_2q\bar q$ for the case when the transition vertex $T\to q\bar
q$ is chosen in the minimal form (12) are shown in Fig. 3$a$.
Form factors decrease noticeably with the increase of radius of the
$q\bar q$ system in the interval 6 Gev$^{-2} \le R_T^2\le 16$
GeV$^{-2}$.
The calculated form factors reveal a strong dependence on the quark
mass. In our calculations we put $m=350$ MeV for the non-strange quark
and $m=500$ MeV for the strange one. One can see that with the increase
of quark mass by 150 MeV the transition form factors fall down, with a
factor 1.5.

In Fig. 3$b$ the form factors
$F^{(H)}_{q\bar q\to\gamma\gamma}(0,0)/Z_{q\bar q}$ are shown for
mesons $1^3P_2 q\bar q$ in case when the vertex $T\to q\bar q$ is taken
in the form (15), that corresponds to a pure $P$-wave.

\subsection{Decays $a_2\to \gamma\gamma$}

The form factor $F^{(H)}_{a_2\to \gamma\gamma}(0,0)$ is equal
to that of the $n\bar n$ component, up to the charge factor:
\be
F^{(H)}_{a_{2}\to \gamma\gamma}(0,0)=\frac{Z_{a_2}}{Z_{n\bar n}}
F^{(H)}_{n\bar n\to \gamma\gamma}(0,0)\ .
\label{2.35}
\ee
Since flavour structure of the $a_2$-meson is fixed, we can
calculate partial $\gamma\gamma$ widths rather reliably.

In Fig. 4 the magnitudes of $\Gamma_{a_2(1320)\to \gamma\gamma}$ are
shown versus $R^2_{a_2(1320)}$ together with experimental data.

Recent measurements of the $\gamma\gamma$ partial width of
the $a_2(1320)$-meson  reported the magnitudes
$\Gamma_{a_2(1320)\to \gamma\gamma}=0.98\pm 0.05 \pm 0.09$ keV
\cite{argus}  and $\Gamma_{a_2(1320)\to \gamma\gamma}=0.96\pm 0.03 \pm
0.13$ keV \cite{l3} (corresponding areas are shown by close hatched
lines in Fig. 4).  Besides, one should take into consideration that the
extraction of the signal $a_2(1320) \to \gamma\gamma$ faces the problem
of a correct account for coherent background. In the analysis
\cite{argus90} it was shown that the measured value of
$\Gamma_{a_2(1320) \to \gamma\gamma}$ can fall down in a factor $\sim
1.5$ due to the interference "signal--background". Therefore,  we
estimate the allowed region for  $\Gamma_{a_2(1320) \to \gamma\gamma}$
 as
$1.12$ keV$\leq \Gamma_{a_2(1320) \to \gamma\gamma}\leq 0.60$ keV (rare
 hatched lines in Fig. 4).

Figure 5$a$ demonstrates a full set of the $a_2\to \gamma\gamma$
widths.
Thick solid line
is drawn for $\Gamma (a_2(1320)\to \gamma\gamma)$
with the vertex given by (9) and (12)
while the
dashed line presents $\Gamma (a_2(1320)\to \gamma\gamma)$ for
the vertex given by  (9) and (15).
The thin solid and dotted lines show correspondingly $\Gamma
(a_2(1660)\to \gamma\gamma)$ for the vertex given by (10) and (12)
and $\Gamma (a_2(\sim 2000)\to \gamma\gamma)$
for the vertex given by  (9) and (16).

When comparing the calculated values with experimental data, one should
keep in mind that quark masses are strictly fixed: as was
 mentioned above, $m=350$ MeV for the non-strange quark
and $m=500$ MeV for strange one. The same mass values have been used in
 \cite{phi} for the calculation of the decays
 $a_0(980) \to  \gamma\gamma$ and  $f_0(980) \to  \gamma\gamma$. Still,
 the two-photon decays of scalar mesons depend comparatively weakly on
 quark masses, while for tensor meson the situation is quite
 opposite:
the decrease of
constituent quark mass by 10\%
results in the increase of the form factor $F^{(H)}_{n\bar
n}(0,0)$  approximately by 10\%, that means a 20\% growth
of the calculated value of $\Gamma_{a_2(1320) \to \gamma\gamma}$ at
fixed $R^2_{a_2(1320)}$. The 10\%-uncertainty in the definition of the
constituent quark mass looks quite reasonable, therefore
the 20\% accuracy of the model prediction for
$\Gamma_{a_2(1320) \to \gamma\gamma}$ should be regarded as quite
normal.

Coming back to the decay $a_2(1320) \to \gamma\gamma$ we  conclude
 that calculated values $\Gamma_{a_2(1320) \to \gamma\gamma}$
 demonstrate rather good agreement with data at $R^2_{a_2(1320)} \la
 12$ GeV$^{-2}$.

\subsection{Decays $f_2(1270) \to \gamma\gamma$ and $f_2(1525) \to
\gamma\gamma$ }

First, consider the decays of mesons belonging to basic $1^3 P_2 q\bar
q$ nonet.
We define
flavour wave functions of $f_2(1270)$ and $f_2(1525)$ as follows:
\bea
f_2(1270):&& \;\;\; \cos\varphi_T\; n\bar n +
\sin \varphi _T\; s\bar s\ ,
\nonumber \\
f_2(1525):&& \;\;\;
-\sin\varphi_T\; n\bar n + \cos \varphi _T \; s\bar s\ .
\label{2.36}
\eea
Then the form factors of the two-photon decays
of $f_2$-mesons read:
\bea
F^{(H)}_{f_2(1270)\to \gamma\gamma}(0,0)&=&  \cos\varphi_T
F^{(H)}_{n\bar n\to \gamma\gamma}(0,0)
 + \sin \varphi _T
F^{(H)}_{s\bar s\to \gamma\gamma}(0,0) , \nonumber \\
F^{(H)}_{f_2(1525)\to \gamma\gamma}(0,0)&=&
-\sin\varphi_T
F^{(H)}_{n\bar n\to \gamma\gamma}(0,0)
 + \cos \varphi _T
F^{(H)}_{s\bar s\to \gamma\gamma}(0,0)\ .
\label{2.37}
\eea

Hadronic decays tell us that $f_2(1270)$ is mainly $n\bar n$ system
while $f_2(1525)$ is $s\bar s$, that is, the mixing angle $\varphi_T$
is small.

Following [26-29]
%\cite{argus}--,l3,argus90,\cite{PDG-00},
we accept  partial
widths as follows: $\Gamma_{f_2(1270) \to \gamma\gamma}=
(2.60\pm{0.25}^{+0.00}_{-0.25})$ keV and
$\Gamma_{f_2(1525) \to \gamma\gamma}=0.097\pm{0.015}^{+0.00}_{-0.25})$
keV. The magnitude of the extracted signal depends on the type of a
model used for the description of the background. For the coherent
background the magnitude of the signal decreases, and the second error
in $\Gamma_{f_2(1270) \to \gamma\gamma}$ and $\Gamma_{f_2(1525) \to
\gamma\gamma}$ is related to the background uncertainties.

 Figure 5$b$ shows the values of $\Gamma_{f_2(1270) \to \gamma\gamma}$
and $\Gamma_{f_2(1525) \to \gamma\gamma}$ calculated under the
assumption that $\varphi_T=0$.

In Fig. 6 the results of the fit to data for
  $\Gamma_{f_2(1270) \to \gamma\gamma}$
and $\Gamma_{f_2(1525) \to \gamma\gamma}$ are shown, without any
fixation of $\varphi_T$.    One can see that there exist two solutions:
 $\varphi_T\simeq 0^\circ$ and $\varphi_T\simeq 25^\circ$, in both
cases $R^2_T \la 10$ GeV$^{-2}$. The rare hatched areas correspond to
the description of data near the low border:
$2.10\,{\rm keV}\le
\Gamma_{f_2(1270) \to \gamma\gamma} \le 2.35\,{\rm keV}$
and
$0.57\,{\rm keV}\le
\Gamma_{f_2(1525) \to \gamma\gamma} \le 0.82\,{\rm keV}$.

\subsection{Two-photon decays of the
$2^3P_2 q\bar q$ and $1^3F_2 q\bar q$ states}

As it follows from \cite{syst}, at 1600--1800 MeV there are two
tensor-isoscalar states -- members of the  $2^3 P_2 q\bar q$
multiplet: they are $f_2(1640)$ and $f_2(1800)$ (presumably $n\bar n$-
and $s\bar s$-dominant states, correspondingly).

In Fig. 3$c$ the form factors of mesons from radial excitation nonet
$2^3P_2 q\bar q$ are shown: the transition vertex $T\to q\bar q$ is
defined by the wave function (10) and spin matrix (12). A small
magnitude of transition form factors  at $R^2_T \ga10$ GeV$^{-2}$ is
due to zero in the wave function (10). The form factor magnitudes
calculated for the transitions  $2^3P_2 q\bar q \to \gamma\gamma$ are
in some way approximate only -- they strongly depend on the details of
the wave functions of  $2^3P_2 q\bar q$-states, and comparatively weak
variation which, for example, does not change  the  mean radius square
$R^2$ may affect the form factor value by 100\%.

Special feature of the two-photon decays of
radial-excitation mesons is that
their partial widths are
considerable smaller than corresponding widths of the basic
mesons:
\bea
&\Gamma_{1^3P_2n\bar n \to \gamma\gamma}&>>
\Gamma_{2^3P_2n\bar n\to \gamma\gamma}\ ,
\nonumber \\
&\Gamma_{1^3P_2s\bar s\to \gamma\gamma}&>>
\Gamma_{2^3P_2s\bar s\to \gamma\gamma}\ .
\label{inequal}
\eea
The inequalities are due to the fact that radial wave function of
$2^3P_2q\bar q$-state contains a zero, therefore the convolution of
wave functions $\psi_{2^3P_2q\bar q}\otimes \psi_{\gamma}$ is
significantly smaller than the convolution
$\psi_{1^3P_2q\bar q}\otimes \psi_{\gamma}$ (wave function of basic
state has no zeros).

In Fig. 5$c$ one can
see partial widths for  $f_2(1640) \to \gamma\gamma$ and $f_2(1800) \to
\gamma\gamma$ calculated under simple hypothesis that $f_2(1640)$ is a
pure $n\bar n$ state, while $f_2(1800)$ is pure $s\bar s$ state. One
should emphasize that the probability for the transition $2^3 P_2 n\bar
n\to \gamma\gamma$  is higher by an order of value than $2^3 P_2 s\bar
s\to \gamma\gamma$. This means that a comparatively small admixture of
the $n\bar n$ component may considerably enhance the width of
$\Gamma_{f_2(1800) \to \gamma\gamma}$, by a factor 2--3 as compared to
what follows from pure $s\bar s$ state.

The verification of Eq. (\ref{inequal}) is of principal meaning from
the point of view of meson quark structure.
Preliminary data of the L3 collaboration \cite{Schegelsky} on the
reaction $\gamma\gamma\to K_s^0K_s^0$ allow one to evaluate the
transition $f_2(1800)\to \gamma\gamma$: Fig. 7 shows the $K_s^0K_s^0$
spectrum where the peaks are distinctly seen which correspond to the
production $f_2(1525)$ and $f_2(1800)$. The description of these peaks
in terms of the Breit--Wigner resonances gives us the following
relation:
\be
\Gamma_{f_2(1800)\to \gamma\gamma}=(0.10\pm 0.05)\,{\rm keV}\cdot
\frac{BR(f_2(1525)\to K\bar K )}{BR(f_2(1800)\to K\bar K )}\ .
\label{l3Gamma}
\ee
Comparing (\ref{l3Gamma}) with partial width
values shown in Fig. 5$c$ proves that the L3 data agree qualitatively
with (\ref{inequal}), provided $BR(f_2(1800)\to K\bar K )\sim
BR(f_2(1525)\to K\bar K )$, that is, the decay channel $f_2(1800)\to
K\bar K $ is not small. A rather large magnitude of the branching
$f_2(1800)\to K\bar K $ looks natural, because the $f_2(1800)$ and
$f_2(1525)$, according to the systematics in the $(n,M^2)$-plane
\cite{syst}, should belong to the same trajectory, so  they both have
rather large $s\bar s$ component. The fact that, according to
(\ref{l3Gamma}), partial width of the decay $f_2(1800)\to \gamma\gamma$
should be greater than $\Gamma_{2^3P_2s\bar s\to \gamma\gamma} \simeq
0.03$ keV may be explained by the (20--30)\% admixture  of  $n\bar n$
component in the $f_2(1800)$.

Figure 3$d$ demonstrates form factors for mesons of the
$1^3F_2 q\bar q$ multiplet: the wave functions are defined by (9),
while the vertex  $T\to q\bar q$ has the form (16).
In Fig. 5$d$ partial widths are shown for the transitions
$1^3 F_2 n\bar n\to \gamma\gamma$ and $1^3 F_2 s\bar s\to \gamma\gamma$
calculated under assumption that masses of these states are of the
order of 2000 MeV \cite{syst}.

\section{Conclusion}

We have calculated the two-photon decays of tensor mesons, members of
the $q\bar q$ multiplets $1^3P_2q\bar q$, $2^3P_2q\bar q$ and
$1^3F_2q\bar q$.

The main goal was to calculate the decays of mesons of basic multiplet
$1^3P_2q\bar q$: $a_2(1320)$, $f_2(1270)$ and $f_2(1525)$.
All calculated partial widths of radiative decays of these mesons,
$a_2(1320)\to \gamma\gamma$, $f_2(1270)\to \gamma\gamma$ and
$f_2(1525)\to \gamma\gamma$, are in reasonable agreement with the
hypothesis about quark-antiquark structure of tensor mesons. In
addition, radial wave functions of $a_2(1320)$, $f_2(1270)$ and
$f_2(1525)$ are close to radial wave functions of $a_0(980)$ and
$f_0(980)$ found in the study of radiative decays
$a_0(980)\to \gamma\gamma$, $f_0(980)\to \gamma\gamma$ and
$\phi(1020)\to \gamma\gamma$ \cite{phi}. The possibility to describe
simultaneously scalar and tensor mesons using approximately equal wave
functions may be considered as a strong argument in favour of the fact
that all these mesons --- tensor $a_2(1320)$, $f_2(1270)$ and
$f_2(1525)$ and scalar $a_0(980)$ and $f_0(980)$ --- are members of the
same $P$-wave $q\bar q$ multiplet.

The mesons of the  first radial excitation, according to \cite{syst},
are $a_2(1660)$, $f_2(1640)$ and $f_2(1800)$. We have calculated
partial $\gamma\gamma$ widths for all these mesons. The comparison with
data of the L3 Collaboration on the reaction $f_2(1800)\to \gamma\gamma$
reveals qualitative agreement.  However, it  should be stressed that
calculated values of partial $\gamma\gamma$ widths of mesons belonging
to the $2^3P_2q\bar q$ multiplet are rather sensitive to details
of the wave function of the $q\bar q$ system.

We have also calculated the
$\gamma\gamma$ width of mesons belonging to the $3^3F_2q\bar q$
multiplet. These mesons are located near 2000 MeV \cite{syst}, and we
may expect them to be a target for studying the reactions
$\gamma\gamma\to$hadrons.

\section*{}
We are grateful to L.G. Dakhno and A.V. Sarantsev for
useful discussions. The paper is suppored by the RFBR grant
$n_0$ 01-02-17861.

\section*{Appendix A. Spin structure of the decay amplitude  $T\to
\gamma\gamma$}

\subsection{The completness of operators $S^{(H)}_{\mu\nu\,\alpha\beta}$}

Here we demonstrate that the convolution of the angular momentum
operators for $L=2$ and $L=4$ with the helicity operator
$S^{(2)}_{\alpha_1 \alpha_2\ ,\ \mu_1\mu_2}$ does not
change the amplitude stucture given by (\ref{2.1}).

The convolution of the helicity $H=2$ operator with that of $L=2$
reads:
$$
X^{(2)}_{\mu_2\beta}(q)S^{(2)}_{\alpha_1 \alpha_2\ ,\ \mu_1\beta}
(p,q)=
\eqno{(A.1)}
$$
$$
=\frac32\left(q^\perp_{\mu_2}
q^\perp_{\beta}-\frac13\, q^2_\perp g^\perp_{\mu_2\beta}\right)
\left(g^{\perp\perp}_{\mu_1\alpha_1} g^{\perp\perp}_{\beta\alpha_2} +
g^{\perp\perp}_{\mu_1\alpha_2} g^{\perp\perp}_{\beta\alpha_1}
-g^{\perp\perp}_{\mu_1\beta}
g^{\perp\perp}_{\alpha_1\alpha_2}\right)=
$$
$$
=-\frac{q^2}{2}
\left(g^{\perp\perp}_{\mu_1\alpha_1} g^{\perp\perp}_{\mu_2\alpha_2} +
g^{\perp\perp}_{\mu_1\alpha_2} g^{\perp\perp}_{\mu_2\alpha_1}
-g^{\perp\perp}_{\mu_1\mu_2}
g^{\perp\perp}_{\alpha_1\alpha_2}\right)=-\frac{q^2}{2}
S^{(2)}_{\alpha_1 \alpha_2\ ,\ \mu_1\mu_2} (p,q).
$$
The convolution of the helicity $H=2$ operator with that of $L=4$
also gives
the term proportional to the $H=2$ operator:
$$
X^{(4)}_{\mu_1\mu_2\beta\lam}(q)
S^{(2)}_{\alpha_1 \alpha_2\ ,\ \nu\lam}(p,q)=
\eqno{(A.2)}
$$
$$
=\frac{35}{8}\left[q_{\mu_1}
q_{\mu_2 }q_{\nu}q_{\lam}
-\frac{q^2}7\left(
g^\perp_{\mu_1\mu_2}q_{\nu}q_{\lam}+
g^\perp_{\mu_1\nu}q_{\mu_2}q_{\lam}+
g^\perp_{\mu_2\nu}q_{\mu_1}q_{\lam}+
g^\perp_{\nu\lam}q_{\mu_1}q_{\mu_2}\right)-
\right .
$$
$$
-\left .
\frac{q^4}{35}\left(
g^\perp_{\mu_1\mu_2}g^\perp_{\nu\lam}+
g^\perp_{\mu_1\nu}g^\perp_{\mu_2\lam}+
g^\perp_{\mu_1\lam}g^\perp_{\mu_2\nu}\right)
\right]
\left(g^{\perp\perp}_{\nu\alpha_1} g^{\perp\perp}_{\lam\alpha_2} +
g^{\perp\perp}_{\nu\alpha_2} g^{\perp\perp}_{\lam\alpha_1}
-g^{\perp\perp}_{\nu\lam}
g^{\perp\perp}_{\alpha_1\alpha_2}\right)=
$$
$$
=\frac{q^4}{4}
\left(g^{\perp\perp}_{\mu_1\alpha_1} g^{\perp\perp}_{\mu_2\alpha_2} +
g^{\perp\perp}_{\mu_1\alpha_2} g^{\perp\perp}_{\mu_2\alpha_1}
-g^{\perp\perp}_{\mu_1\mu_2}
g^{\perp\perp}_{\alpha_1\alpha_2}\right)=\frac{q^4}{4}
S^{(2)}_{\alpha_1 \alpha_2\ ,\ \mu_1\mu_2}(p,q).
$$
So, we see that the both convolutions, the $H=2$ operator with
$L=2$ and $L=4$, give the terms proportional to
$S^{(2)}_{\alpha_1 \alpha_2\ ,\ \mu_1\mu_2} (p,q)$.

\subsection{The operators $S^{(H)}_{\mu\nu\,\alpha\beta}$ and standard
helicity technique}

To demonstrate  the connection of the introduced
operators with standard helicity technique, consider as an
example the transition $\gamma\gamma\to 2^{++}-
{\rm resonance}\to\pi^{0}\pi^{0} $.
Using the momenta $q_1$, $q_2$ for photons and
$k_1$, $k_2$ for mesons, one has for relative momenta
and photon polarization vectors
$$
q=\frac12(q_1-q_2)=(0,0,0,q_z),
\qquad k=\frac12(k_1-k_2) =(0,k_x,k_y,k_z),
\eqno{(A.3)}
$$
$$
\epsilon=(0,\epsilon_x,\epsilon_y,0)=(0,\cos{\phi},
\sin{\phi},0).
$$
In the helicity basis
$$
\epsilon=(0,\epsilon_+,\epsilon_-,0),
\eqno{(A.4)}
$$
$$
\epsilon_+=-\frac1{\sqrt{2}}(\epsilon_x+i\epsilon_y),\qquad
\epsilon_-=\frac1{\sqrt{2}}(\epsilon_x-i\epsilon_y)
$$
The spin-dependent part of the amplitude with $H=0$ reads:
$$
\epsilon^{(1)}_{\alpha}\epsilon^{(2)}_{\beta}S^{(0)}_{\alpha\beta}
X^{(2)}_{\mu\nu}(q)X^{(2)}_{\mu\nu}(k)=
\frac{9}{4}q^{2}k^{2}(\cos^2 \theta-\frac13)
(\epsilon^{(1)}_{+}\epsilon^{(2)}_{+}+\epsilon^{(1)}_{-}\epsilon^{(2)}_{-}).
\eqno{(A.5)}
$$
For $H=2$ one has
$$
\epsilon^{(1)}_{\alpha}\epsilon^{(2)}_{\beta}S^{(2)}_{\alpha\beta\,\mu\nu}
X^{(2)}_{\mu\nu}(k).
\eqno{(A.6)}
$$
Different components are written as follows:
$$
\epsilon^{(1)}_{+}\epsilon^{(2)}_{+}S^{(2)}_{+ + \mu\nu}
X^{(2)}_{\mu\nu}(k)=0,
\eqno{(A.7)}
$$
$$
\epsilon^{(1)}_{+}\epsilon^{(2)}_{-}S^{(2)}_{+ - \mu\nu}
X^{(2)}_{\mu\nu}(k)=\frac32k^2\sin^2\theta
%\frac32k^2(1+2\sin^2{\phi}-2i\sin{\phi}\cos{\phi})=
\left(1+2i\sin{\phi}\,e^{i\phi}\right),
$$
$$
\epsilon^{(1)}_{-}\epsilon^{(2)}_{+}S^{(2)}_{- + \mu\nu}
X^{(2)}_{\mu\nu}(k)=\frac32k^2\sin^2\theta
%(1+2\sin^2{\phi}+2i\sin{\phi}\cos{\phi})=
\left(1-2i\sin{\phi}\,e^{-i\phi}\right) ,
$$
$$
\epsilon^{(1)}_{-}\epsilon^{(2)}_{-}S^{(2)}_{- - \mu\nu}
X^{(2)}_{\mu\nu}(k)=0.
$$

\subsection{Operators $S^{(H)}_{\mu\nu\,\alpha\beta}$  and analytical
properties of vertex function}

The operators  $S^{(H)}_{\mu\nu\,\alpha\beta}$  are expressed through
metric tensors $g^{\perp\perp}_{\alpha\beta}$ which work in the space
perpendicular to the reaction plane $T\to \gamma\gamma$. This metric tensor
has the structure as follows:
$$
g^{\perp\perp}_{\alpha\beta}=g_{\alpha\beta}-p_\alpha p_\beta/m^2_T+
4q_\alpha q_\beta/m_T^2\ .
\eqno{A.8} $$

The presence of factors $1/m_T^2$ may evoke the question about the behaviour
of the form factor amplitude at $m_T^2=0$. Of course, this is a far remote
point for the reactions under  consideration (the lowest tensor resonance is
at $m_T \simeq 1600$ GeV$^2$). Still, this problem sounds principal, so
consider it in details.

When treating the reaction "composite system $\to \gamma\gamma$" one should
distinguish it from the transition "constituents $\to \gamma\gamma$", in
this case $q\bar q \to \gamma\gamma$. These processes though related to each
other are different: the amplitude for the process
"composite system $\to \gamma\gamma$" (or $T\to \gamma\gamma$) is determined
by the residue  in the amplitude pole $q\bar q\to \gamma\gamma$ at $s=m^2_T$,
where $s$ is the invariant energy squared of the $q\bar q $ system. Both
amplitudes should satisfy the requirement of gauge invariance (this
requirement for the transition $T\to \gamma\gamma$ is imposed by the
operators $S^{(H)}_{\mu\nu\,\alpha\beta}$), but they have different
analytical properties. In particular, the threshold theorems appropriate for
$q\bar q\to \gamma\gamma$ do not take place in the vertex function, i.e.
for the residue in the pole $s=m_T^2$ the threshold theorems are realised
by the interplay of pole and non-pole terms.  Due to this fact we must keep
 the mass of the tensor meson $m_T^2$ in Eq.  (A.8) as fixed value.

The problem of interrelation of gauge invariance and analyticity in the
spectral integration technique has been discussed in detail in
\cite{eta,spin-1} for the transitions $q\bar q \to \gamma q\bar q$,
$NN\to \gamma NN$, $NN\gamma \to NN\gamma$ and corresponding vertex functions
describing composite systems (mesons and deuteron).

An opposite point of view according to which the threshold theorems should
work for vertex functions was outspoken in \cite{Acha}.

\section*{Appendix B. Spin factors for pure $q\bar q(L=1)$ and
$q\bar q(L=3)$ states}

At small $Q^2$ the spin factor
$S_{L=1}^{(H=0)}(P\ ^2=s,P\ '^2=s',-Q^2)$ for pure $(L=1)$-state
reads:
$$
S_{L=1}^{(0)}(s,s',-Q^2)=
\eqno{(B.1)}
$$
$$
=\frac{32\Sigma^2Q^4}{(\Delta^2+4\Sigma
Q^2)^2}\left[2m^2
\left(4m^2-\Sigma\right)-
\frac{m\left(4m^2-\Sigma\right)^2}{2m+\sqrt{\Sigma}}\right]+
$$
$$
+\frac{4\Sigma Q^2\Delta^2}{(\Delta^2+4\Sigma
Q^2)^2}\left[\left(32m^4+8m^2\Sigma-3\Sigma^2\right)-
\frac{4m\left(16m^4-\Sigma^2\right)}{2m+\sqrt{\Sigma}}\right]+
$$
$$
+\frac{4m^2\Delta^4}{(\Delta^2+4\Sigma
Q^2)^2}\left[\left(4m^2+3\Sigma\right)-
\frac{4m\left(2m^2+\Sigma\right)}{2m+\sqrt{\Sigma}}\right]
$$
Recall that we use the
notations $\Sigma =(s+s')/2$ and $\Delta =s-s' $.

For $H=2$ one has:
$$
S_{L=1}^{(2)}(s,s',-Q^2)=
\eqno{(B.2)}
$$
$$
=\frac{8\Sigma^2 Q^4}{(\Delta^2+4\Sigma
Q^2)^2}\left[\left(-16m^4+\Sigma^2\right)+
\frac{2m\left(4m^2-\Sigma\right)^2}{2m+\sqrt{\Sigma}}\right]+
$$
$$
+\frac{4\Sigma Q^2\Delta^2}{(\Delta^2+4\Sigma
Q^2)^2}\left[\left(-16m^4-4m^2\Sigma+\Sigma^2\right)+
\frac{8m^3\left(4m^2-\Sigma\right)}{2m+\sqrt{\Sigma}}\right]+
$$
$$
+\frac{4m^2\Delta^4}{(\Delta^2+4\Sigma
Q^2)^2}\left[\left(-2m^2-\Sigma\right)+
\frac{4m^3}{2m+\sqrt{\Sigma}}\right].
$$
Correspondingly, the spin factors
$S_{L=3}^{(H)}(s,s',-Q^2)$
for pure $(L=3)$-state are as follows:
$$
S_{L=3}^{(0)}(s,s',-Q^2)=
\eqno{(B.3)}
$$
$$
=\frac{48\Sigma^2 Q^4}{5(\Delta^2+4\Sigma Q^2)^2}\left[
2m^2\left(4m^2-\Sigma\right)^2-
\frac{m\left(4m^2-\Sigma\right)^3}{2m+\sqrt{\Sigma}}\right]+
$$
$$
+\frac{12\Sigma Q^2\Delta^2}{5(\Delta^2+4\Sigma Q^2)^2}\left[
\left(16m^4+4m^2\Sigma+\Sigma\right)\left(4m^2-\Sigma\right)-
\frac{2m\left(4m^2+\Sigma\right)\left(4m^2-\Sigma\right)^2}
{2m+\sqrt{\Sigma}}\right]+
$$
$$
+\frac{12m^2\Delta^4}{5(\Delta^2+4\Sigma Q^2)^2}\left[
\left(8m^4+4m^2\Sigma+\Sigma^2\right)-
\frac{2m\left(4m^2-\Sigma\right)\left(2m^2+\Sigma\right)}
{2m+\sqrt{\Sigma}}\right],
$$
and
$$
S_{L=3}^{(2)}(s,s',-Q^2)=
\eqno{(B.4)}
$$
$$
=\frac{4\Sigma^2 Q^4}{5(\Delta^2+4\Sigma Q^2)^2}\left[
\left(-6m^2+\Sigma\right)\left(4m^2-\Sigma\right)^2+
\frac{3m\left(4m^2-\Sigma\right)^3}
{2m+\sqrt{\Sigma}}\right]+
$$
$$
+\frac{2\Sigma Q^2\Delta^2}{5(\Delta^2+4\Sigma Q^2)^2}\left[
\left(-24m^4+4m^2\Sigma-\Sigma^2\right)\left(4m^2-\Sigma\right)+
\frac{12m^3\left(4m^2-\Sigma\right)^2}
{2m+\sqrt{\Sigma}}\right]+
$$
$$
+\frac{2m^2\Delta^4}{5(\Delta^2+4\Sigma Q^2)^2}\left[
\left(-12m^4+2m^2\Sigma-\Sigma^2\right)+
\frac{6m^3\left(4m^2-\Sigma\right)}
{2m+\sqrt{\Sigma}}\right] \ .
$$

\section*{Appendix C. Transition form factors
$F^{(H)}_{T(L)\to \gamma \gamma}(0,0)$}

Spectral integral for the form factor
$F^{(H)}_{T(L)\to \gamma \gamma}(0,0)$
with pure $(L=1)$ or $(L=3)$ wave in the vertex $T\to q\bar q$
reads as follows:
$$
F^{(H)}_{T(L)\to \gamma \gamma}(0,0)=
\frac{Z^{(q\bar q)}\sqrt{N_c}}{16\pi}
\int \limits_{4m^2}^\infty
\frac{ds}{\pi} \psi_T(s)\psi_{\gamma}(s)I^{(H)}_{L}(s) \ ,
\label{}
\eqno{(C.1)}
$$
where for $(L=1)$ one has
$$
I_{L=1}^{(0)}(s)=-2\sqrt{s\left(s-4m^2\right)}
\left(12m^2+3\right)
+4m^2\left(4m^2+3s\right)
     \ln\frac{s+\sqrt{s\left(s-4m^2\right)}}
             {s-\sqrt{s\left(s-4m^2\right)}}+
$$
$$
+\frac{16m^3}{2m+\sqrt{s}}\left[
3\sqrt{s\left(s-4m^2\right)}
-\left(2m^2+s\right)
\ln\frac{s+\sqrt{s\left(s-4m^2\right)}}
        {s-\sqrt{s\left(s-4m^2\right)}}\right]
\eqno{(C.2)}
$$
and
$$
I_{L=1}^{(2)}(s)=\frac{4\sqrt{s\left(s-4m^2\right)}}{3}
\left(5m^2+s\right)
-4m^2\left(2m^2+s\right)
        \ln\frac{s+\sqrt{s\left(s-4m^2\right)}}
                {s-\sqrt{s\left(s-4m^2\right)}}+
$$
$$
+\frac{4m}{2m+\sqrt{s}}\left[
-\frac{\sqrt{s\left(s-4m^2\right)}}{3}
\left(10m^2-s\right)
+4m^4\ln\frac{s+\sqrt{s\left(s-4m^2\right)}}
             {s-\sqrt{s\left(s-4m^2\right)}}\right]\ .
\eqno{(C.3)}
$$
Analogously, for the $q\bar q(L=3)$-wave ($1^3F_2q\bar q$ multiplet),
we have
$$
I_{L=3}^{(0)}(s)=-\frac{2\sqrt{s\left(s-4m^2\right)}}{5}
\left(72m^4+8m^2s+s^2\right)+
\eqno{(C.4)}
$$
$$
\,\,\,+\frac{12}{5}m^2\left(8m^4+4m^2s+s^2\right)
        \ln\frac{s+\sqrt{s\left(s-4m^2\right)}}
                {s-\sqrt{s\left(s-4m^2\right)}}+
$$
$$
+\frac{24m^3\left(s-4m^2\right)}{5(2m+\sqrt{s})}\left[
-3\sqrt{s\left(s-4m^2\right)}
+\left(2m^2+s\right)\ln\frac{s+\sqrt{s\left(s-4m^2\right)}}
                            {s-\sqrt{s\left(s-4m^2\right)}}\right]
$$
and
$$
I_{L=3}^{(2)}(s)=\frac{2\sqrt{s\left(s-4m^2\right)}}{15}
\left(30m^4-4m^2s+s^2\right)-
\eqno{(C.5)}
$$
$$
\,\,\,\,\,\,\,\,-\frac{2}{5}m^2\left(12m^4-2m^2s+s^2\right)
        \ln\frac{s+\sqrt{s\left(s-4m^2\right)}}
                {s-\sqrt{s\left(s-4m^2\right)}}+
$$
$$
+\frac{m\left(s-4m^2\right)}{5(2m+\sqrt{s})}\left[
\sqrt{s\left(s-4m^2\right)}
\left(10m^2-s\right)
-12m^4\ln\frac{s+\sqrt{s\left(s-4m^2\right)}}
              {s-\sqrt{s\left(s-4m^2\right)}}\right]\ .
$$

\newpage
\begin{figure}
%fig.1
\vspace{3cm}
\centerline{\epsfig{file=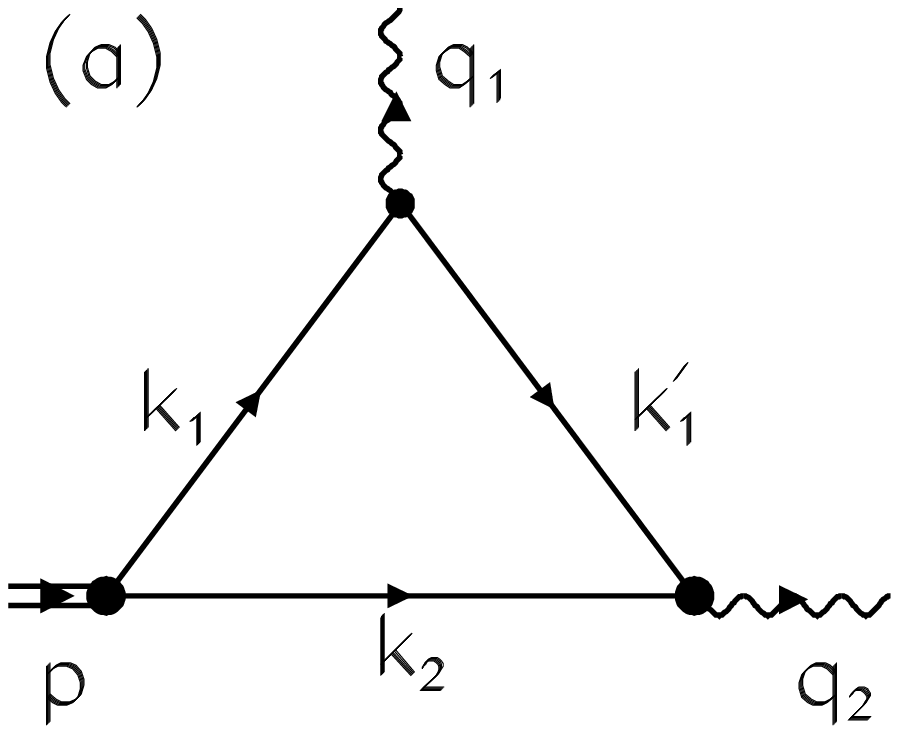,width=4.0cm}\hspace{1cm}
            \epsfig{file=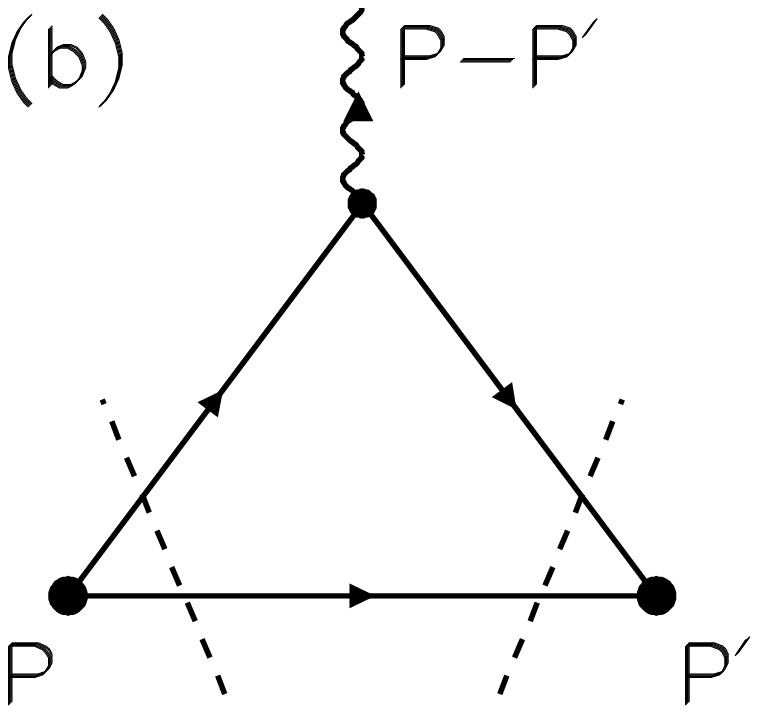,width=4.0cm}\hspace{1cm}
            \epsfig{file=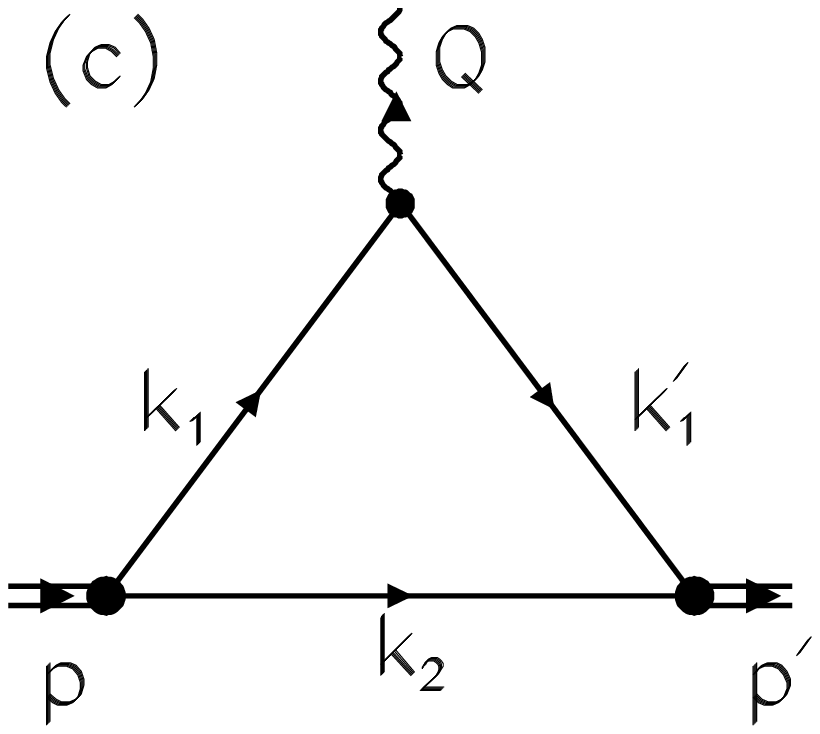,width=4.0cm}}
\caption{(a) Triangle diagram for the transition form factor
$T\to\gamma (q^2_1)\gamma (q^2_2)$; (b) diagram for the double spectral
representation over $P^2=s$ and $P'^2=s'$; the
intermediate-state particles
are mass-on-shell, the cuttings of the diagram are
shown by dashed lines; (c) triangle diagram for the meson charge form
factor.}
\end{figure}

\begin{figure}
%fig.2
\centerline{\epsfig{file=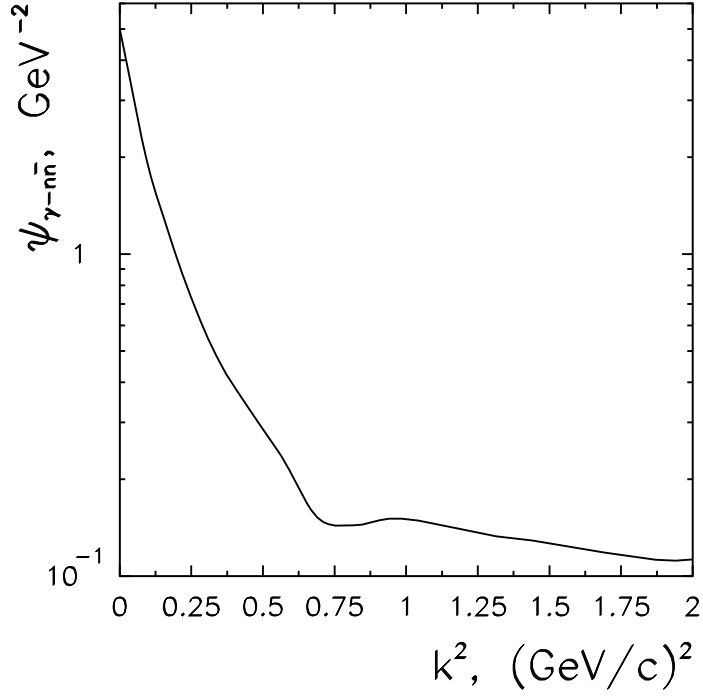,width=10cm}}
\caption{Photon wave function for non-strange quarks,
$\psi_{\gamma\to n\bar n}(k^2)=g_\gamma(k^2)/(k^2+m^2)$, where
$k^2=s/4-m^2$; the wave function for the $s\bar s$
component is equal to
$\psi_{\gamma\to s\bar s}(k^2)=g_\gamma(k^2)/(k^2+m^2_s)$ where $m_s$
is the constituent $s$-quark mass.}
\end{figure}

\begin{figure}
%fig.3
\centerline{\epsfig{file=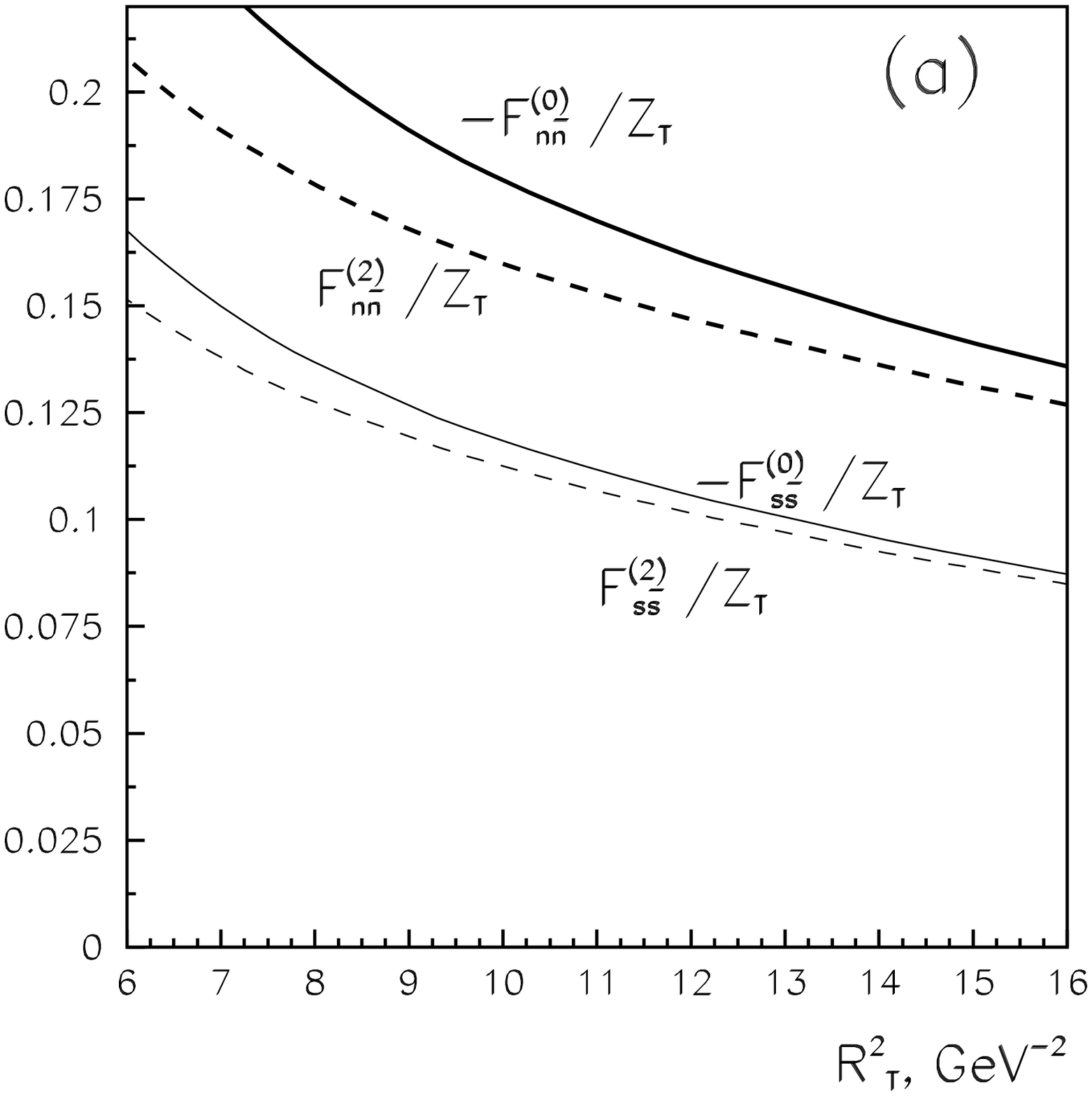,width=7cm}
            \epsfig{file=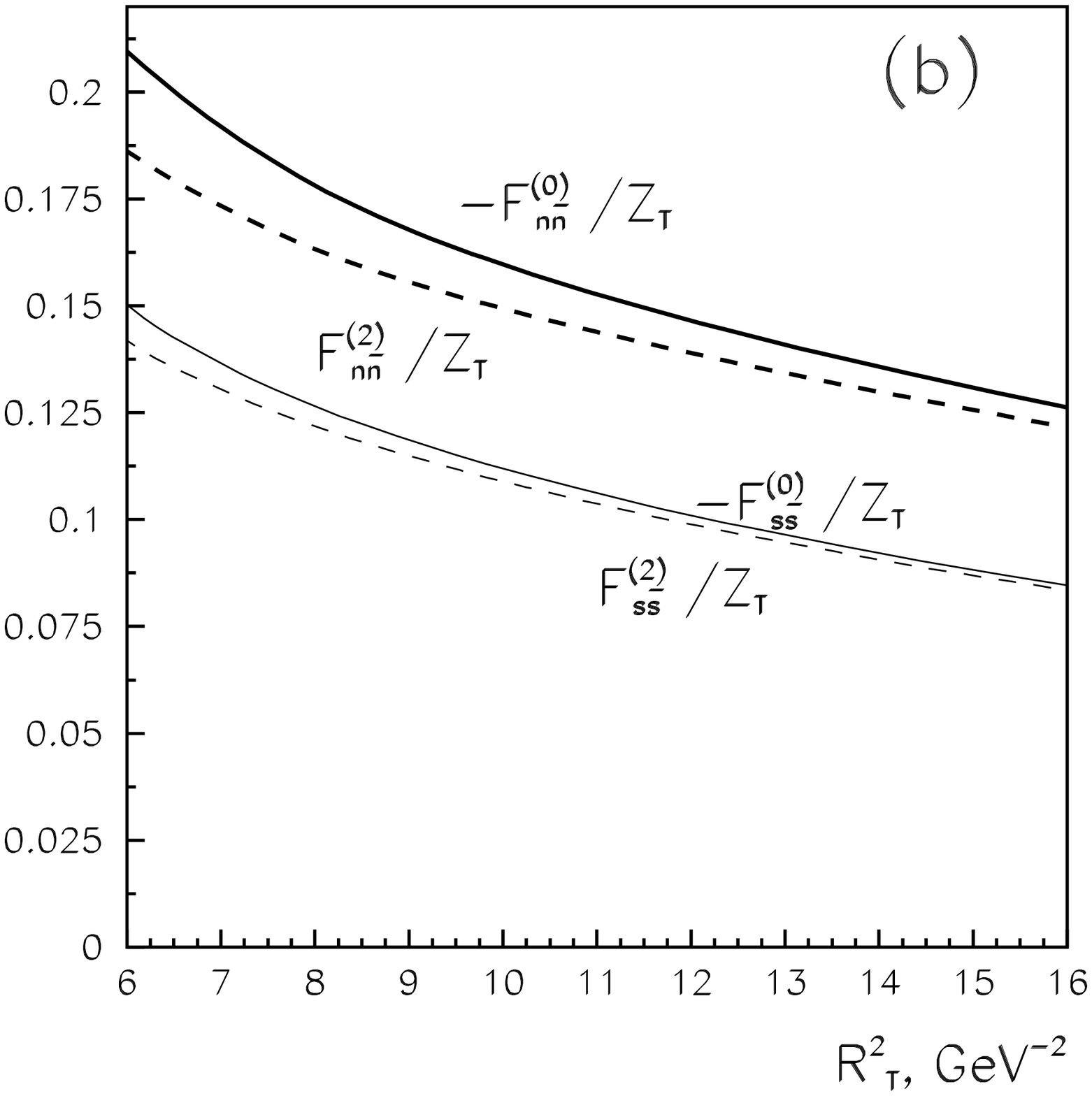,width=7cm}}
\centerline{\epsfig{file=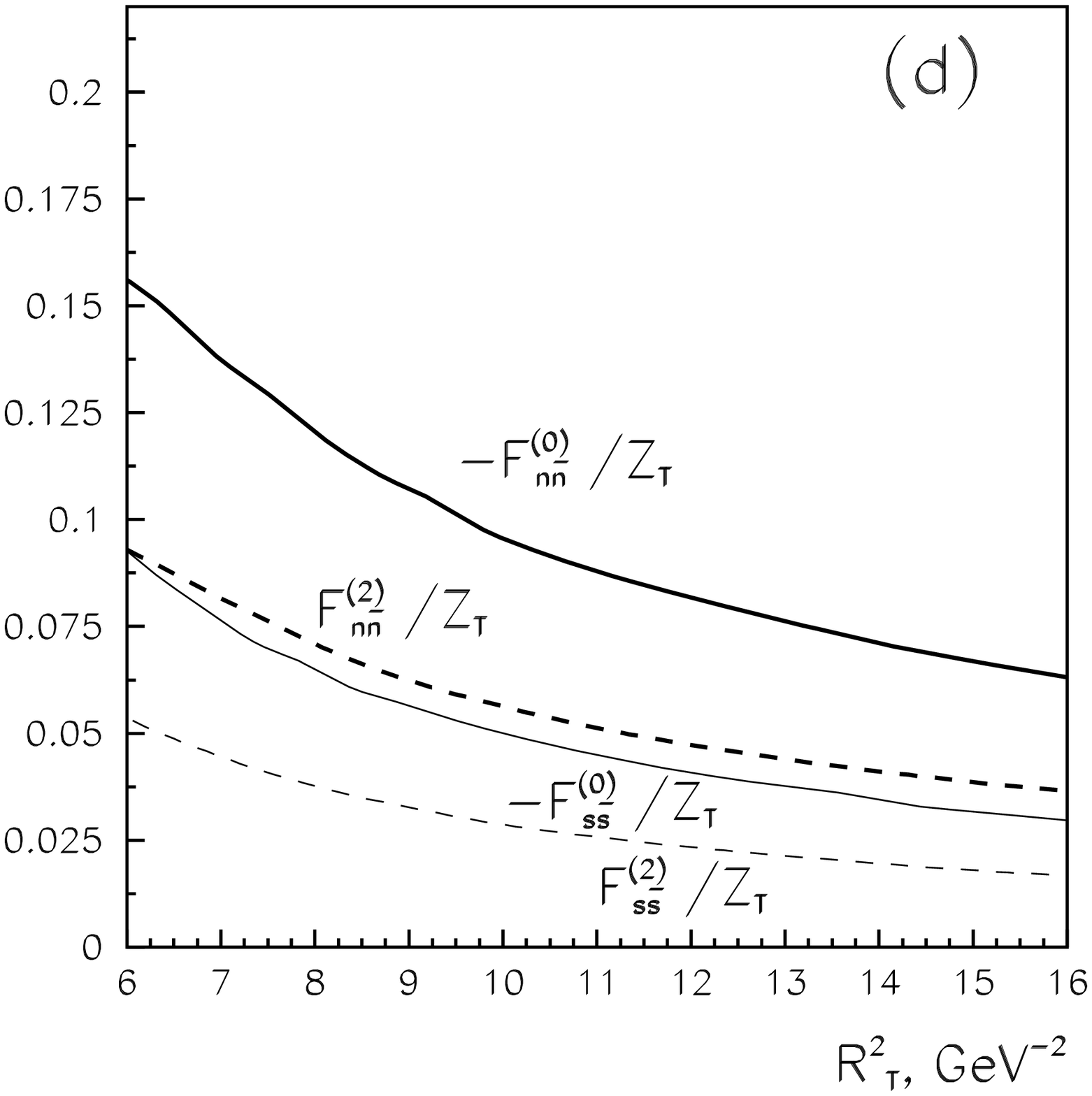,width=7cm}
            \epsfig{file=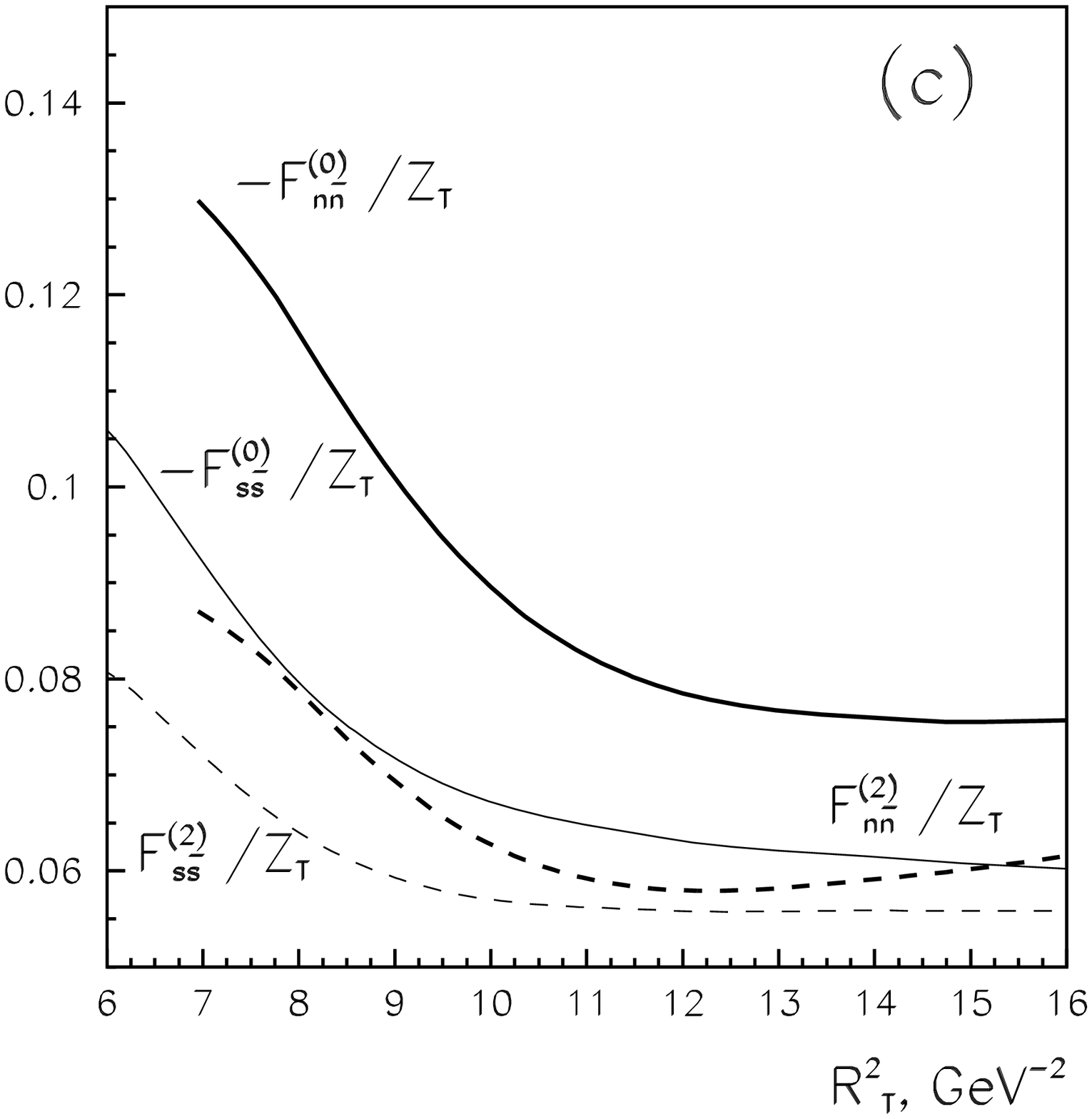,width=7cm}}
\caption{Transition form factors $T\to \gamma\gamma$ (see (17)
or (24)) for the non-strange ($n\bar n$) and strange ($s\bar s$)
quarks versus mean tensor meson radius squared, $R^2_T$.  (a)
$F^{(H)}_{q\bar q}(0,0)$ for $1^3P_2 q\bar q$ state with minimal
vertex, Eqs. (9) and (12); (b) the same as  Fig. 3a but for the
vertex determined by (15);
(c) $F^{(H)}_{q\bar q}(0,0)$ for $2^3P_2 q\bar q$ state
with the vertex given by (10) and (12);
d) $F^{(H)}_{q\bar q}(0,0)$ for $1^3F_2 q\bar
q$ state with vertex given by  (9) and (16).}
\end{figure}

\begin{figure}
%fig.4
\centerline{\epsfig{file=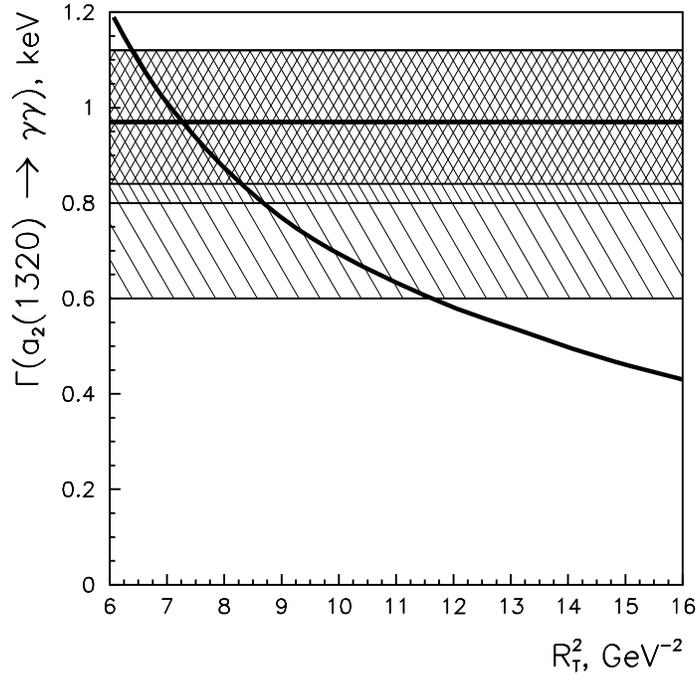,width=10cm}}
\caption{Partial width $\Gamma (a_2(1320)\to \gamma\gamma)$ (thick
solid line) versus the data (hatched areas, see section 3.3) as a
function of the mean meson radius squared, $R^2_T$, for the
vertex determined by Eqs. (9) and (12).  }
\end{figure}

\begin{figure}
%fig.5
\centerline{\epsfig{file=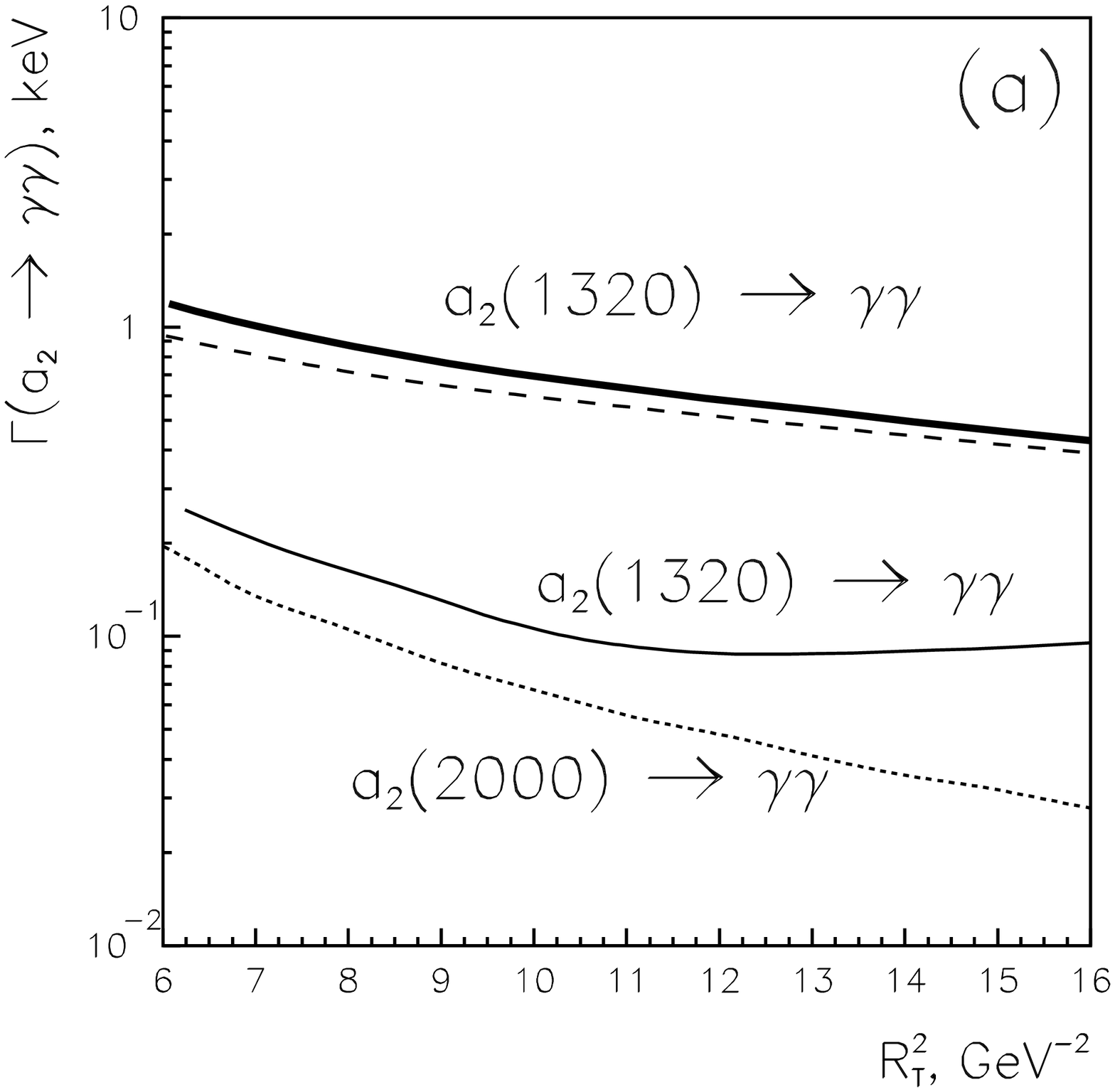,width=7cm}
            \epsfig{file=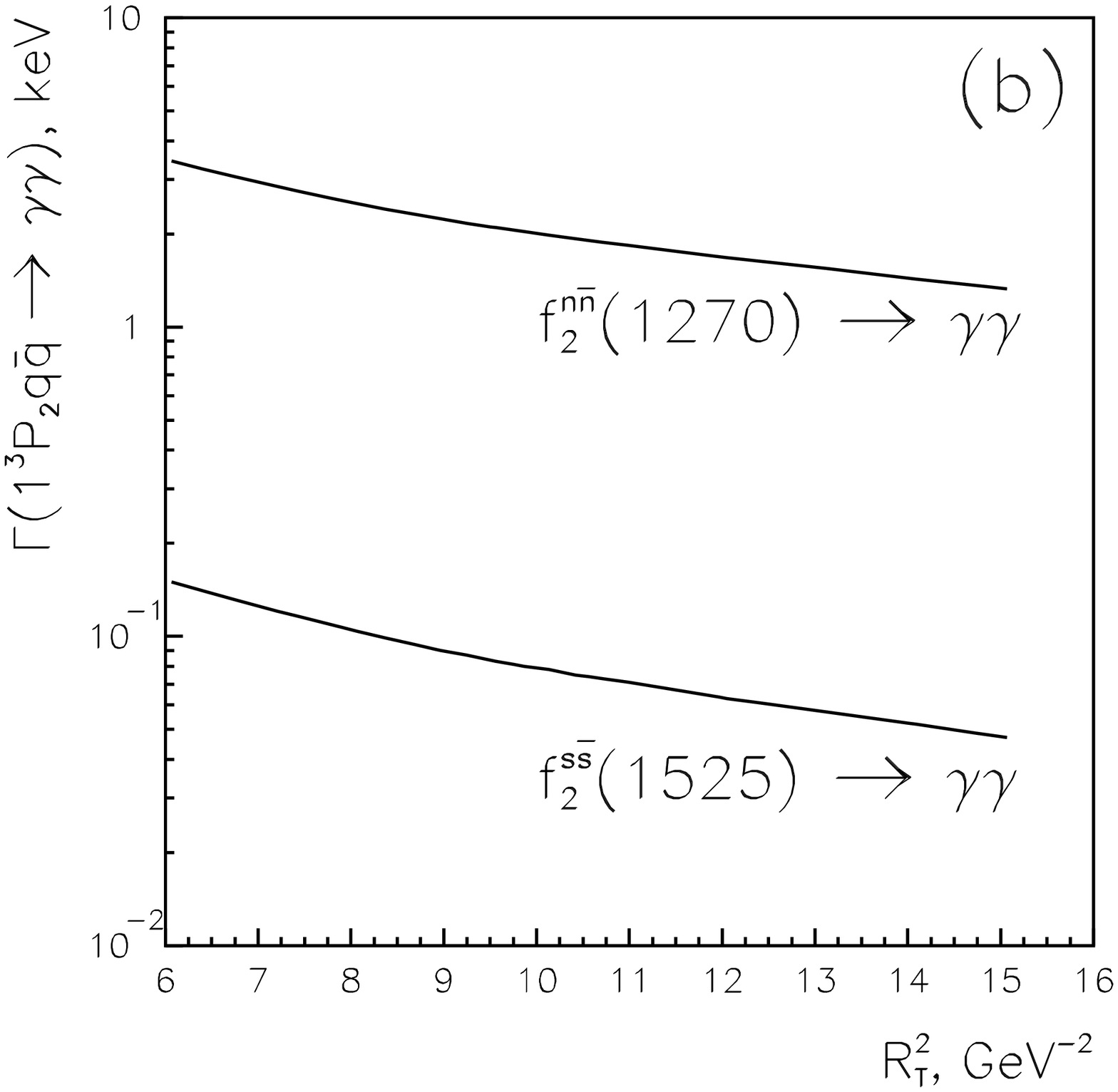,width=7cm}}
\centerline{\epsfig{file=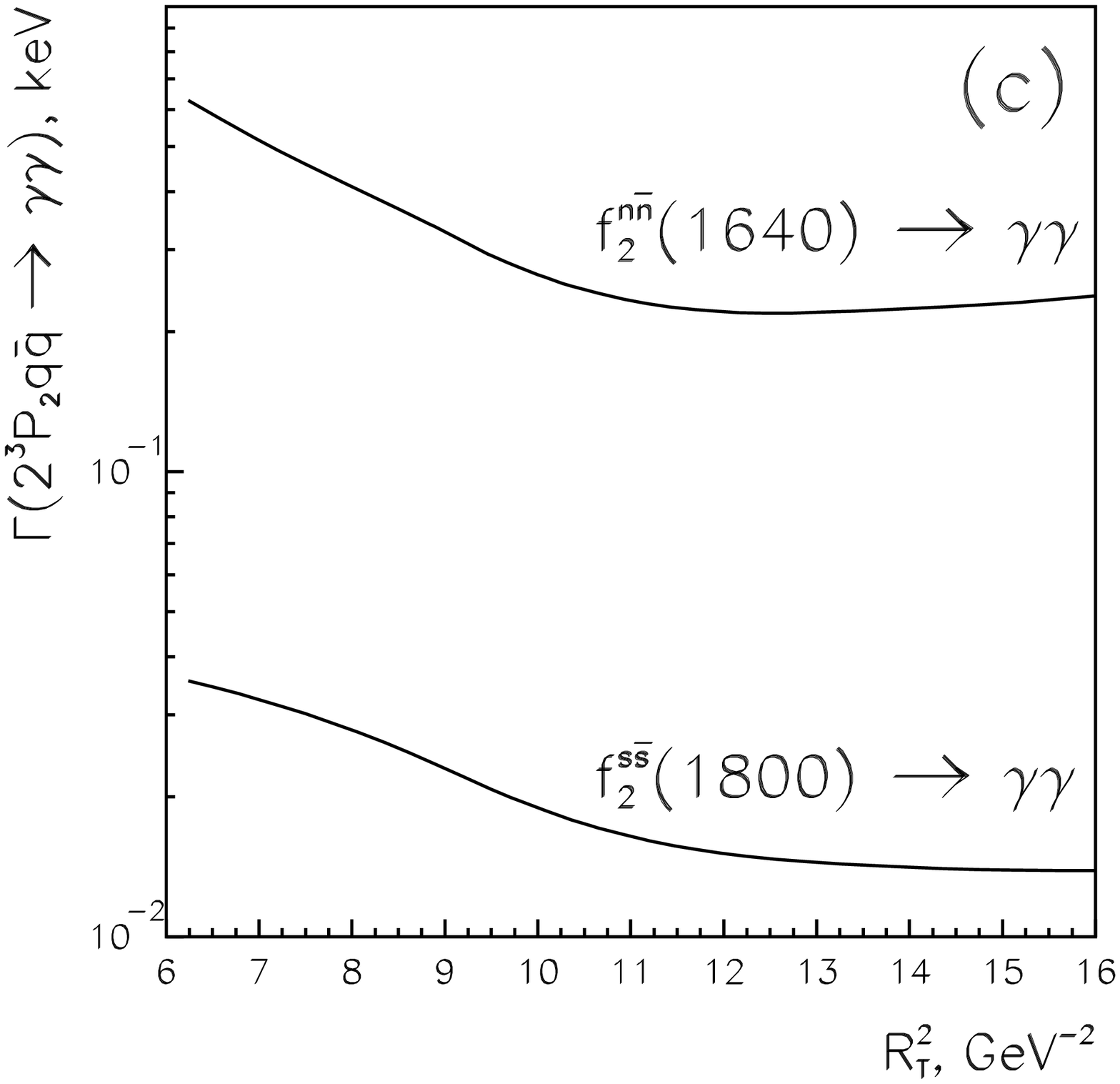,width=7cm}
            \epsfig{file=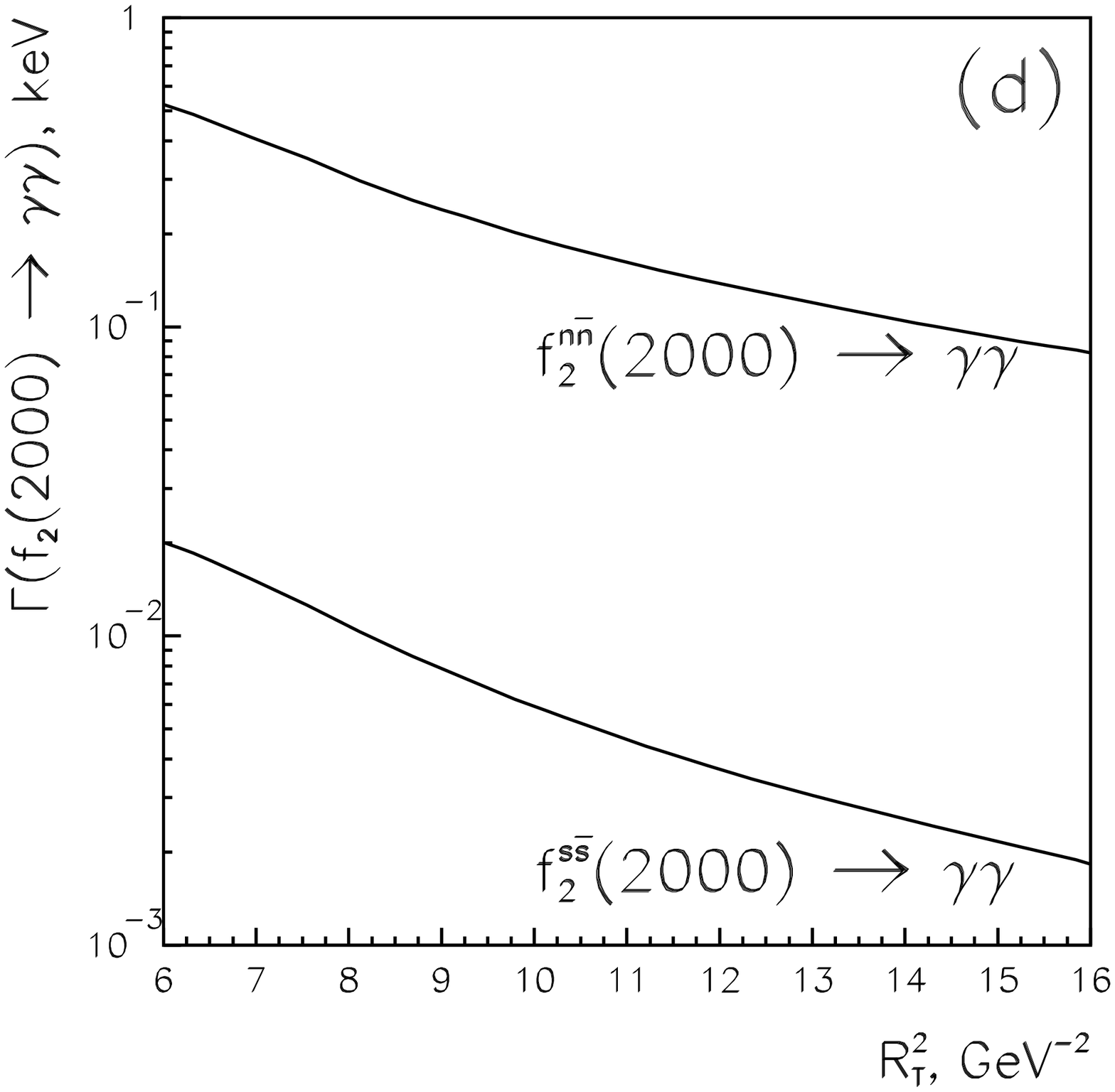,width=7cm}}
\caption{Partial widths for $a_2$ and $f_2$ mesons
versus mean tensor-meson radius squared, $R^2_T$.
(a) Thick solid line: $\Gamma (a_2(1320)\to \gamma\gamma)$
for the vertex given by  (9) and (12);
dashed line: $\Gamma (a_2(1320)\to \gamma\gamma)$
for the vertex given by  (9) and (15);
dotted line: $\Gamma (a_2(\sim 2000)\to \gamma\gamma)$
for the vertex given by  (9) and (16);
thin solid line: $\Gamma (a_2(1660)\to \gamma\gamma)$
for the vertex given by  (10) and (12).
(b) Partial widths for isoscalar mesons of the basic
$1^3P_3q\bar q$ nonet: for $f_2(1525)$ meson supposing it is
either pure $s\bar s$ state
and for $f_2(1270)$ under assumption it is pure $n\bar n$.
(c) Partial width for $f_2(1800)$ meson supposing it is
the pure $2^3P_2 s\bar s$ state and for $f_2(1640)$  considered
as a pure $2^3P_2 n\bar n$ state.
(d) Partial widths for $1^3F_2 s\bar
s$ and $1^3F_2 n\bar n$ states with mass $\sim 2000$ MeV.}
\end{figure}

\begin{figure}
%fig.6
\centerline{\epsfig{file=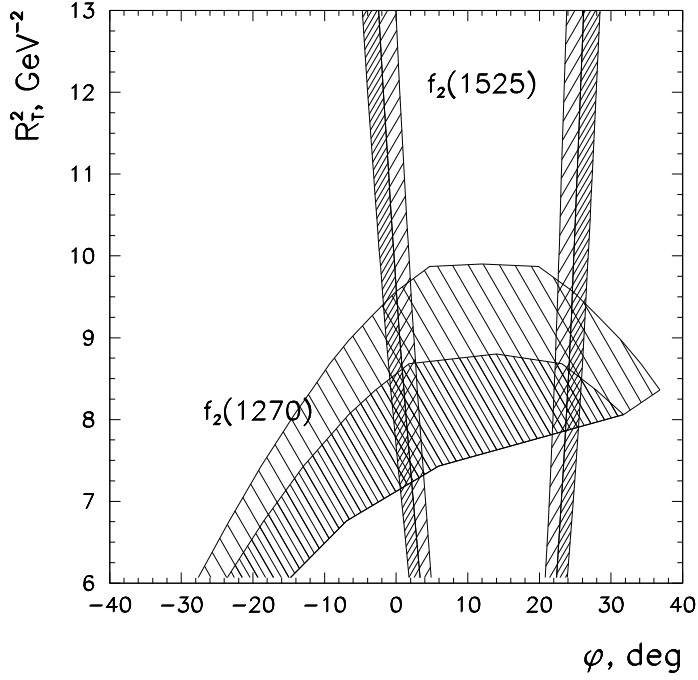,width=10cm}}
\caption{The $(R^2_T,\varphi)$-plot, where $\varphi$ is mixing angle
for the flavour components
$f_2(1270)=n\bar n\cos\varphi  +
 s\bar s\sin \varphi $ and $ f_2(1525)=
- n\bar n\sin\varphi +  s\bar s\cos \varphi $,
with hatched areas which show the regions allowed by data for decays
$f_2(1270)\to \gamma\gamma$ and $f_2(1525)\to \gamma\gamma$.}
\end{figure}

\begin{figure}
%fig.7
\centerline{\epsfig{file=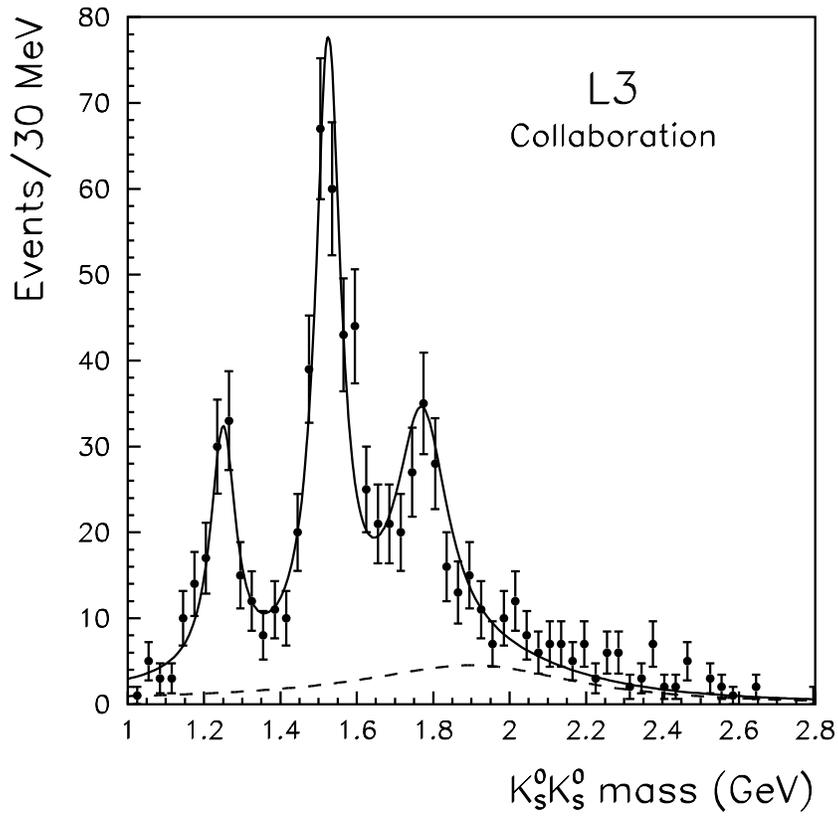,width=12cm}}
\caption{The $K^0_sK^0_s$ mass spectrum in
$\gamma\gamma \to K^0_sK^0_s$ [30] with production $f_2(1525)$
and  $f_2(1800)$.}
\end{figure}

\end{document}